\documentclass[journal,twoside,web]{ieeecolor}

\usepackage{generic}
\usepackage{cite}
\usepackage{amsmath,amssymb,amsfonts}
\usepackage{algorithmic}
\usepackage{graphicx}
\usepackage{textcomp}
\usepackage[outdir=./]{epstopdf}
\usepackage{caption}
\usepackage{subcaption}

\usepackage{comment}
\specialcomment{cmtPai}{\begingroup\sffamily\scriptsize\color{red}}{\endgroup}

\newcommand{\FIG}[1]{Fig.~\ref{#1}}
\newcommand{\EQN}[1]{Eqn.~\eqref{#1}}

\newcommand{\id}[1]{\ensuremath{\mathit{#1}}}

\def\BibTeX{{\rm B\kern-.05em{\sc i\kern-.025em b}\kern-.08em
    T\kern-.1667em\lower.7ex\hbox{E}\kern-.125emX}}
\begin{document}
\title{Analysis of Cardiovascular Changes Caused by Epileptic Seizures in Human Photoplethysmogram Signal}
\author{Seyede Mahya Safavi, Ninaz Valisharifabad, Robert Sabino, Hsinchung Chen, Ali HeydariGorji, Demi Tran, Jack Lin, Beth Lopour,
Pai H. Chou
}

\maketitle

\begin{abstract}
Objectives: This study examines human Photoplethysmogram
(PPG) along with Electrocardiogram (ECG) signals to study cardiac autonomic imbalance in epileptic seizures. The significance and the prevalence of changes in PPG morphological parameters have been investigated to find common patterns among subjects. Alterations in cardiovascular parameters measured by PPG/ECG signals are used to train a neural network based on LSTM for automatic seizure detection. Methods: Electroencephalogram (EEG), ECG, and PPG signals from 12 different subjects ( 8 males;4 females;age 34.3$\pm$ 13.8) were recorded including 57 seizures and 101 hours of inter-ictal data. 12 PPG features significantly changing due to epileptic seizures were extracted and normalized based on a proposed z-score metric. 7 feature are heart rate variability related and 5 features hemodynamic related. Results: A consistent pattern of ictal change was observed for all the features across the subjects/seziures. The proposed seizure detector is subject independent and works for both nocturnal and diurnal seizures. With an average of 0.52 false alarms per hour, positive predictive value of $43\%$ and sensitivity of $92\%$, the new proposed hemodynamic based seizure detector shows improvement over the the heart rate variability based detector. Conclusion: The cardiac autonomic imbalance due to seizure manifests itself in variations of peripheral hemodynamics measured by PPG signal, suggesting vasoconstriction in limbs.  These variations can be used on a consumer seizure detecting devices with optical sensors for seizure detection. Significance: The stereotyped pattern is common among all the subjects which can help understand the mechanism of cardiac autonomic imbalance induced by epileptic seizures.

\end{abstract}


\section{Introduction}
 \label{sec:introduction}
\IEEEPARstart{E}{pilepsy} is a chronic disease affecting more than
50
 million people world wide \cite{demog}. In majority of
patients, the seizures are controlled using medication; however,
one-third of
 the patients still do not respond to treatments and
continue to have
 seizures. Although epilepsy is considered a
neurological disorder, several studies have reported autonomic
imbalance and cardiovascular dysfunction during ictal and postictal
period \cite{heart of epi}. In fact, Sudden Unexpected Death in
Epilepsy (SUDEP) has been attributed to cardiovascular and pulmonary
dysfunction induced by uncontrolled seizures \cite{SUDEP}.
 Recently,
epilepsy is considered as a set of coexisting comorbids,
 where
Electroencephalography (EEG) abnormal waves can be preceded by other
extracerebral
 manifestations \cite{autonomic}.\\
 \indent A
personal diary of seizures plays a key role in clinical
 diagnosis
and research. However, keeping a report of the seizures is
 not
trivial, especially for patients who have seizures during sleep
\cite{record}. In addition, since the patient may lose consciousness
during a seizure, timely intervention by a caregiver may be
required.
 In order to have an accurate record of seizures and
identify the overlapping symptoms, new seizure detecting devices have
adopted multimodal data fusion techniques incorporating small sensors
such as
 electrodermal\cite{EDA,EDA-seizure}, accelorometer \cite{ACC}, electromyogram \cite{EMG}, and electrocardiogram (ECG)
\cite{devices,automatic_detection,ECG_nocturnal,ECG_singlelead,ECG_feedback}. The use of multimodal sensors
facilitates automatic seizure detection without using bulky and
uncomfortable EEG caps. However, the performance of these techniques are dependent on the seizure type \cite{automatic_detection}. Specifically electrodermal, accelorometer, and electromyogram sensors are used for generalized tonic-clonic seizures and their performance degrades for partial seizures.\\
 \indent
Ictal tachycardia is the most prevalent autonomic imbalance
manifestation observed in 82\% of seizures \cite{ictal-tachycardia}.
Ictal tachycardia
 happens as a result of increase in sympathetic
tone. However, in less
 than 5\% of cases, parasympathetic activity
can also predominate,
 leading to bradycardia
\cite{ictal-bradycardia}. Studies suggest a consistent and
stereotyped
 progression of autonomic dynamics leading to heart rate
variability (HRV) during and even before the
 seizure onset \cite{model}.
The HRV due to autonomic imbalance has been
exploited for seizure prediction a few seconds before the seizure
onset \cite{prediction}. In general, the HRV-based seizure detectors perform poorly in terms of high false alarm rates and low sensitives compared to EEG-based detectors \cite{devices}. Besides heart rate variability, other less
prevalent cardiac conduction
 abnormalities are inconsistent with
respect to age, laterization,
 and seizure foci. Abnormalities such
as atrial fibrillation, bundle
 branch block, atrial premature
depolarization, asystole, and
 ST-segment elevation have also been
reported in total of less than
 14\% of investigated cases and were not used for seizure detection
\cite{heart of epi}.\\
\indent In addition to cardiac conduction
abnormalities, the excess
 release of catecholamines during seizure
will influence the vascular
 function and hemodynamics.
Catecholamines such as norepinephrine are types of
neurotransmitters released to act as
 anticonvulsant in ictal phase
\cite{norepinephrine}. These
 neurotransmitters are responsible for a
series of autonomic responses
 such as blood flow manipulation. The
blood perfusion is controlled by
 constriction and dilation of the
vessels performed by muscle cells
 present in the blood vessel wall.
These cells are abundant with alpha
 adrenegric receptors that are
the targets of neurotransmitters such as
 norepinephrine. The release
of norepinephrine calls off the blood from
 non-vital organs leading
to a reduction in skin perfusion in limbs. The skin
 vasoconstriction
due to release of catecholamines can be measured by
photoplethysmogram (PPG) \cite{ppg_hemody}. PPG is an optical sensor
composed of a light emitting diode and a photo detector. The
reflection of the light from skin captures the volumetric changes of
the blood pulse. Conventionally, PPG signal was used to measure
blood
 oxygenation level and heart rate. Recent studies have shown
the
 morphology of PPG signal contains valuable information about
the
 cardiovascular function, autonomic nervous system (ANS) and its
related
 hemodynamics \cite{arousal,ppg_hemody,stressPPG}. The blood pressure and vascular compliance affect the morphological features of a PPG pulse. In addition, the HRV features conventionally derived from ECG signal can still be derived using PPG signal and the results of seizure detection is comparable with ECG signal \cite{MPDIsensor}.\\
 In this study, an Empatica E4 wristband is used to
capture PPG signals from 12 subjects while they are being monitored in
UCI Medical Center. In addition to PPG, the EEG and the ECG signals are also recorded.
The seizures are marked by a neurologist and the PPG/ECG data is
segmented to have at least one seizure in every four hours of data. Next the ictal changes in morphological features of PPG signal related to ANS is extracted. In addition to HRV-related features which were conventionally derived and used in ECG-based seizure detectors, other ANS-related morphological features of PPG are being analyzed for the first time.
The extracted morphological features from PPG signal is analyzed for
common patterns pre- and post-seizures and show consistent pattern of change in majority of subjects. We proposed a recurrent neural network based on Long Short-Term Memory (LSTM). LSTM architecture can learn long-term temporal dependencies which helps detecting the progression of features leading to seizure. Despite state of art HRV-based detectors which are trained specifically for each subject, our proposed seizure detector is subject-independent and trained generally for all subjects.
The proposed seizure detector keeps the sensitivity up to 92\% while reducing the false positive rate to 0.52 times per hour while which is $70\%$ improvement.\\

\indent This paper is organized as follows: Section \ref{sec:material}
explains the extracted PPG features and their physiological
interpretation. Section \ref{sec:data_analysis} describes the
procedure in which the data was recorded and the details of the
subjects' information. In addition, the results of analysis and observations are brought in Section \ref{sec:data_analysis}. Section \ref{sec:ML} brings the proposed seizure detection algorithm. Section \ref{sec:conclusion} concludes the paper.


\section{Methods and Material}
\label{sec:material}
\begin{figure}[!t]
\centerline{\includegraphics[trim=90 20 50 20,width=180pt]{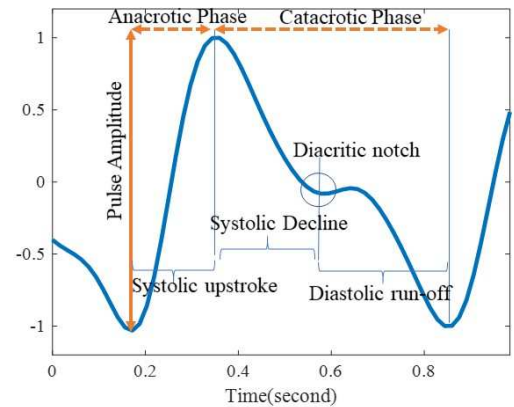}}
\caption{PPG pulse.}
\label{fig:ppg}
\end{figure}

PPG is an optical sensor capturing the skin blood flow using a light
emitting diode (LED) and a photo detector. The LED illuminates the
skin, and the backscattered light from the surface of the skin changes
with respect to blood volume pulse and the blood color. The captured
signal represents the arterial pressure as the blood is ejected from
the left ventricle, circulates through the vessels, and finally goes
back to the right atrium. The PPG signal has a distinct shape that can be
separated into anacrotic and catacrotic phases as shown in
\FIG{fig:ppg}.

The anacrotic phase consists of the systolic upstroke corresponding to the to acceleration of the aortic blood flow as a result of left ventricular blood ejection. The ascending slope depends on several parameters such as the left ventricular ejection pressure, arterial peripheral resistance, and the arterial compliance and elasticity \cite{arterial_waveform}. The upstroke reaches the maximum arterial pressure point called the systolic point. People with less arterial compliance and elasticity tend to have higher systolic peaks.

The catacrotic phase consists of the systolic decline, the dicrotic
notch, and the diastolic run-off. The systolic decline happens when
the left ventricular contraction is about to end. The efflux of blood
from arteries to veins back to the heart is faster than the influx of
the blood from the heart left ventricle. The dicrotic notch happens
when the aortic valve closes. There is a sudden increase in arterial
pressure after the dicrotic notch. In fact, after the closure of the
aortic valve, the blood volume has less space to move and it can only
travel toward the peripheral arteries, causing a secondary peak in
arterial pressure in diastolic phase due to resistance of peripheral
arteries. The dicrotic notch is more slurred in people with a higher
arterial compliance. Finally, there is a gradual drop in the arterial
pressure in diastolic phase as the blood goes back to the right atrium.
\begin{figure}[!t]
\centerline{\includegraphics[trim=90 20 50 20,width=180pt]{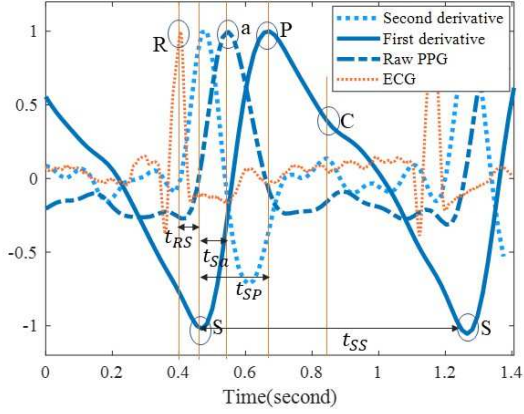}}
\caption{PPG features.}
\label{fig:derppg}
\end{figure}
\subsection{Data Analysis}
In this section the investigated morphological features derived from ECG and PPG data is described along with their physiological interpretation. The temporal and spectral features related to HRV, i.e. $\text{HR}$, $\text{SDNN}$, $\text{RMSSD}$, $\text{NN50}$, $\text{LF}$, $\text{HF}$, and $\text{LF/HF}$, were conventionally derived from ECG signal and are called HRV-related features hence forth. There are 5 other features related to hemodynamics and vascular compliance derived from PPG morphology, i.e. ${t}_{\text{NMV}}$, ${t}_{\text{NCT}}$, $\text{PA}$, $\text{PCA1}$, and $\text{PTT}$. For brevity, these 5 features are called hemodynamic-related features.
\subsubsection{Heart Rate (HR)}
Heart rate is estimated using the time between troughs of consecutive
PPG pulses. Representing the pulse width by $t_{\text{SS}}$ as shown in
\FIG{fig:derppg}, heart rate is defined as the reciprocal of the pulse
width as:
\begin{equation}\label{hr}
	\text{HR}=\frac{1}{t_{\text{SS}}}.
\end{equation}
In addition to heart rate, there are other temporal and spectral parameters extended from heart rate which have been conventionally used as bio-markers of ANS dynamics and can be extracted from PPG signal as follows \cite{MPDIsensor}:
\begin{itemize}
\item \textbf{SDNN}:
SDNN is the standard deviation of pulse widths and is defined as:
\begin{equation}\label{sdnn}
	\text{SDNN}=\mathbb{E}\{(\text{t}_{SS}-\mu_{\text{t}_{SS}})^2\},
\end{equation}
where $\mathbb{E}\{.\}$ is the expectation denotation and $\mu_{t_{SS}}$ is the mean value of the pulse widths.
\item \textbf{RMSSD}:
RMSSD is the root mean square of the expected value of squared differences of successive pulse widths and is defined as:
\begin{equation}\label{rmssd}
	\text{RMSSD}=\sqrt{\mathbb{E}\{(\text{t}_{SS}[i+1]-\text{t}_{SS}[i])^2\}},
\end{equation}
where $t_{SS}[i]$ represents the $i^{th}$ pulse width.
\item \textbf{NN50}:
NN50 is the number of successive pulses with more than $50 msec$ difference in their widths. 
\item {$\boldsymbol{\text{LF}_{norm}}$}:
$LF_{norm}$ is the power spectral density of heart rate in low frequency band (0.04-0.15 Hz) and normalized to the total spectral power of heart rate.
\item $\boldsymbol{\text{HF}_{norm}}$:
$\text{HF}_{norm}$ is the power spectral density of heart rate in high frequency band (0.15-0.4 Hz) and normalized to the total spectral power of heart rate.
\item $\boldsymbol{\text{LF/HF}}$:
$\text{LF/HF}$ is the ratio of power spectral density of heart rate between the low and high frequency band.
\end{itemize}
\subsubsection{Pulse Amplitude (PA)}
 PA is defined as the height of the PPG signal, which is measured by the vertical distance between the diastolic trough to the systolic peak of the next pulse as shown in \FIG{fig:ppg}. PA is directly related to cardiac volume stroke, vascular distensibility, and vascular resistance \cite{ppg_morphology}. In case of hypovolemia and dehydration, the left ventricular volume stroke is small and the PA goes low. In addition, in case of vasoconstriction in peripheral arteries where the vascular resistance of peripheral arteries goes high, a reduction in the PA is observed \cite{PTT,diameter}. PA is also related to arterial compliance and elasticity. Compliance is the ability of blood vessel wall to distend in response to changes in blood pressure. People with lower arterial compliance tend to have a higher PA \cite{ppg_morphology}.

\subsubsection{Normalized Crest Time ($t_{\text{NCT}}$)} The crest time is the
time between the start point of systolic upstroke (denoted by S in
\FIG{fig:derppg}) and the systolic peak (denoted by P in \FIG{fig:derppg}) and is
represented by $t_{\text{SP}}$. Disregarding the fact that $t_{\text{SP}}$ is
expected to reduce due to tachycardia, normalizing the crest time with respect to
pulse width reveals an increase in ictal and post-ictal phase. In fact, regardless
of ictal reduction in crest time, the increase in normalized crest time
($T_{\text{NCT}}$) yields information about vasoconstriction and vascular resistance in limbs during seizure. The normalized crest time is defined as:
\begin{equation}\label{NCT}
	t_{\text{NCT}}=\frac{t_{\text{SP}}}{t_{\text{SS}}}.
\end{equation}
\subsubsection{Normalized Pulse Transmit Time ($\text{PTT}$)}
Pulse transmit time is the time it takes from the onset of the left ventricle
depolarization to the time the blood reaches the peripheral arteries and is
measured as the interval between the R peak in ECG signal and the next PPG trough
as shown by $t_{\text{RS}}$ in \FIG{fig:derppg} \cite{PEP}. The pulse transmit time is related to pulse wave velocity or the speed of blood flow in arteries. The pulse wave velocity itself is a function of blood density, arterial dimension properties such as vessel thickness and arterial diameter, and blood pressure \cite{PTT}. The normalized pulse transmit time is defined as follows:
\begin{equation}\label{NPTT}
	\text{PTT}=\frac{t_{\text{RS}}}{t_{\text{SS}}}.
\end{equation}

\subsubsection{Maximum Velocity and its Normalized Time ($t_{\text{NMV}}$ )} The point of maximum upstroke slope in systolic phase denoted by peak $a$ in the first derivative waveform represents the maximum velocity in PPG pulse \FIG{fig:derppg}. This maximum slope depends on the blood viscosity, arterial pressure and vascular resistance. In addition the normalized maximum slope time is defined as:
\begin{equation}\label{NST}
	t_{\text{NMV}}=\frac{t_{Sa}}{t_{\text{SS}}}.
\end{equation}
\subsubsection{Principle Component from the Second Derivative of PPG (PCA1)} The second derivative of the PPG  pulse shape (SDP) is a
measure of blood flow acceleration in vessels. The fiducial points of the second derivative
waveform are related to arterial stiffness and arterial pressure
\cite{elghandi2012}. In this study, the raw PPG data is segmented into PPG pulses.
Each PPG pulse starts from the diastolic trough and ends in the diastolic trough
of the next pulse as shown in \FIG{fig:derppg}. After calculating the second derivative of each pulse shape, the dominant shape of
the second derivative of the pulses in inter-ictal phase (baseline) is
derived using principle component analysis (PCA). All the PPG pulses occurring
before 15 minutes prior to the seizure onset and after 5 minutes post-seizure
offset are considered as baseline \cite{prediction}. The shapes of the SDP in
ictal phases are compared to the principle components of the
baseline using subspace projection. In order to be able to apply PCA to the second derivative of PPG pulse shape, the data samples are interpolated such that the PPG pulses have the same number of samples denoted by
$N$.
 Let us represent the $k^{\text{th}}$ pulse as a $N \times 1$ vector $\boldsymbol{b}_k$ and its second derivative as the baseline acceleration vector $\boldsymbol{b}''_k$. Stacking all the baseline acceleration vectors, the $N\times K$ baseline acceleration matrix $\boldsymbol{B}$ is formed as:
\begin{equation}\label{B}
  \boldsymbol{B}=[\boldsymbol{b}''_1, \boldsymbol{b}''_2, ...,\boldsymbol{b}''_K]_{N\times K},
\end{equation}
where $K$ is the total number of baseline pulses.
An eigenvalue decomposition of its covariance is as follows;
\begin{equation}\label{eigen}
  \frac{1}{K}\boldsymbol{B}\boldsymbol{B}^T=\boldsymbol{\Psi}\boldsymbol{\Lambda}\boldsymbol{\Psi}^T,
\end{equation}
where $\boldsymbol{B}^T$ is the transpose of $\boldsymbol{B}$ and $\boldsymbol{\Psi}=[\boldsymbol{\psi}_1,\boldsymbol{\psi}_2,...,\boldsymbol{\psi}_N]$ is the $N\times N$ matrix of eigenvectors  with $\boldsymbol{\psi}_i$ being the $i^{th}$ eigenvector. $\boldsymbol{\Lambda}$ is an $N \times N$ diagonal matrix of eigenvalues. The eigenvectors corresponding to the major eigenvalues contain the dominant shape of the baseline acceleration waveform. In order to compare the ictal SDP shape with baseline, we adopted a subspace projection approach as explained in the following. Let us represent the ictal PPG pulses and their second derivatives by $N \times 1$ vectors ${\boldsymbol{b}}_\ell$ and ${\boldsymbol{b}}''_{\ell}$, where $\ell$ is the index of ictal pulse. In order to quantify the deviation of each acceleration waveform from the baseline, the subspace projection of ${\boldsymbol{b}}''_{\ell}$ with the subspace spanned by the principle components in \EQN{eigen} is derived as follows:
\begin{equation}\label{correlation}
  \text{PCA1}[\ell]=\frac{{\boldsymbol{\psi}_1}^T{\boldsymbol{b}}''_{\ell}{{\boldsymbol{b}}''_{\ell}}^T\boldsymbol{\psi}_1}{{{\boldsymbol{b}}''_{\ell}}^T{\boldsymbol{b}}''_{\ell}},
\end{equation}
where $\boldsymbol{\psi}_1$ is the first principle component of baseline SDP and $\ell$ is the index of the ictal pulse.

 \section{Data Collection and Analysis }
 \label{sec:data_analysis}
\subsection{Data Collection}
A total of 12 subjects (8 males and 4 females with a mean age of
33.64 $\pm$ 13.3) were recruited among the patients with refractory
epilepsy who underwent long-term extra-cranial monitoring admitted to
University of California Irvine Medical Center (UCIMC). The informed
consents were obtained before the start of the monitoring as required
by the Institutional Review Boards of University of California,
Irvine. The 20-channel surface EEG data and one-lead ECG data (Lead II) were
continuously recorded
as a standard of care procedure with sampling
rate of 500~Hz (The Nihon Kohden JE-921, paired with the QI-123A LAN converter). Our study also involved collection of PPG data recorded simultaneously by Empatica E4 with 64~Hz of sampling
frequency. The PPG data collected the blood pulse from the left ankle
of the subjects. The subjects were laying on a bed while being
monitored to reduce the motion artifact. The seizure onset time and
foci were extracted by a clinical neurophysiologist based on the
revision of the EEG data. A total of 60 seizures were recorded where 3 seizures were withdrawn from the analysis due to extensive noise in PPG data.
Table \ref{table:subjects} represents the subjects' clinical information
including the age and the seizure onset zone.
\begin{table*}[t]
\caption{Subjects' Clinical Characteristics}
\centering
\begin{tabular}{c c c c c c c c}
\hline
\hline
Patient&
Sex&
Age& Seizure type & Origin & Number of seizures & Medication & hour:min   \\
\hline
1 &Male &27 &Partial &Unknown &1 & Clonazepam, Depakote &2:0' \\
2 &Female &26 &Generalized &- &1 & Topiramate, Lamictal&1:30'\\
3 &Male &27 &Partial &Left Frontal, Left Temporal &17 & Keppra, Vimpat &19:0'\\
4 &Male &23 &Partial &Right Frontal, Right Temporal &7 & Trileptal&23:21' \\
5 &Male &58 &Partial &Left Temporal &1 & Keppra, Levetricateram &5:0'\\
6 &Female &50 &Partial &Right Temporal &12 & Clonazepam, Levetricateram&16:0'\\
7 &Male &25 &Partial &Right Frontal, &5 & Trileptal, Zonisamide&10:41' \\
8 &Female &33 &Partial & Left Temporal, Left Parietal &1 & Phenytoin &4:0'\\
9 &Male &27 &Partial & Right Temporal, &1 & Lamictal, Clonazepam, Depakote, Mirtazapine &4:0'\\
10 &Male &25 &Partial & Left Occipital, Left Parietal &1 & Zonisamide&4:0' \\
11 &Male &59 &Partial & Left Temporal &9 & Keppra, Topiramate &21:0'\\
12 &Female &21 &Partial & Left Frontal &1 & Zonisamide, Phenytoin&2:0' \\
\hline
\end{tabular}
\label{table:subjects}
\end{table*}
\begin{table*}[t]
\caption{Collected Seizures}
\centering
\begin{tabular}{c c c c c c c c c c c c c c c c c c c c c}
\hline
\hline
 Seizure Number & S1 & S2 & S3 & S4 & S5& S6 & S7 & S8& S9 & S10 &S11 & S12& S13 & S14 & S15 & S16 & S17  &S18 & S19\\
  \hline
  Patient number & P1 & P2& P3 & P3& P3 & P3 & P3 & P3 & P3 & P3 & P3 & P3 & P3 & P3 & P3 & P3 & P3 & P3 & P3 \\
  Duration(sec) &110& 70 & 60 & 78 &117 & 113 & 42 & 46 & 56 &40 & 63 &175 &37 &102 &81 & 297 &19  & 85 & 59 \\
  \hline
  \hline \\
  Seizure Number &  S20& S21 & S22 & S23 & S24 & S25 & S26 & S27& S28 & S29 & S30& S31 & S32 & S33 & S34 & S35 &S36 &S37 & S38
  \\
  \hline
    Patient number  & P4 &P4& P4  &P4& P4 &P4&P4 &P5 &P6 & P6 & P6  & P6 & P6 & P6 & P6 &P6& P6 & P6 & P6\\
  Duration(sec) & 39 & 84 & 93 & 71 & 98 & 104 & 61 & 79 & 94 & 91 & 201 & 156 & 49 & 42 & 35 & 35 & 36 & 47 &48 \\
  \hline
  \hline \\
 Seizure Number & S39 &S40& S41 & S42 &S43 &S44 &S45 &S46 & S47 & S48 & S49 & S50 & S51 &S52 & S53 & S54 &S55 &S56 &S57\\
\hline
 Patient number  & P6&P7& P7 &P7&P7 &P7  & P8  & P9 & P10 & P11 & P11& P11 & P11& P11 & P11& P11 & P11 & P11 &P12\\
  Duration(sec) & 59 & 36& 31 & 39 & 32 & 39& 28 & 265 & 13 & 54& 44 &48 &71 &45 & 37&61 &29 &38 &72 \\
  \hline
  \hline\\

\end{tabular}
\label{table:seizures}
\end{table*}

\subsubsection{Timing Synchronization}
\begin{figure}
\centerline{\includegraphics[trim=90 20 50 20,width=140pt]{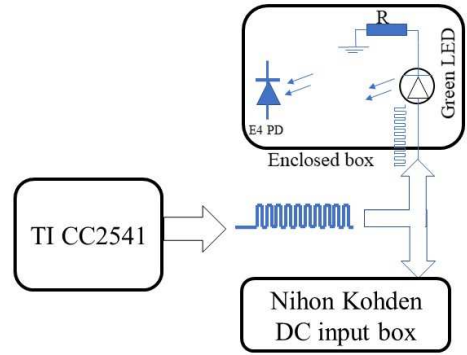}}
\caption{Synchronization apparatus}
\label{fig:synch}
\end{figure}

Since the EEG/ECG data and the PPG data are recorded from separate devices,
	the time offset and the drift noise present in the clocks of the two devices
	will cause timing error in recording events. In order to synchronize the events
	in Nihon Kohden EEG recording machine and the Empatica E4 PPG recording device,
	we designed an apparatus that generates time stamps twice during each recording
	session. The time stamp is a sequence of 10 square pulses with a pulse width of
	1 second and duty cycle of 50\%.  The Nihon Kohden JE-921 has an analog input
	box that enables the collection of an external analog signal \cite{NK}. The
	designed synchronizer feeds the analog input port of Nihon Kohden with the
	sequence square pulses (5 volts of pulse amplitude). At the same time, the E4 device is placed in front of a green LED which is fed with the same sequence of pulses. As a result, the LED blinks with a frequency of 1 Hz and duty cycle of 50\%.  The photo detector embedded in E4 captures the LED flashes as the time stamp. The time stamp is generated two times, once initially before attaching the E4 to the subject's  left ankle and once again at the end after detaching it from the subject's body. \FIG{fig:synch} shows the block diagram of the synchronization apparatus. Once the recording session is finished, the timing is tuned by maximally aligning the time stamps captured by E4 photo detector and Nihon Kohden analog input box.
\label{subsub:analysis}
\begin{figure*}

	\centering
	\begin{subfigure}{0.3\textwidth} 
		\includegraphics[trim=0 10 0 20,width=\textwidth]{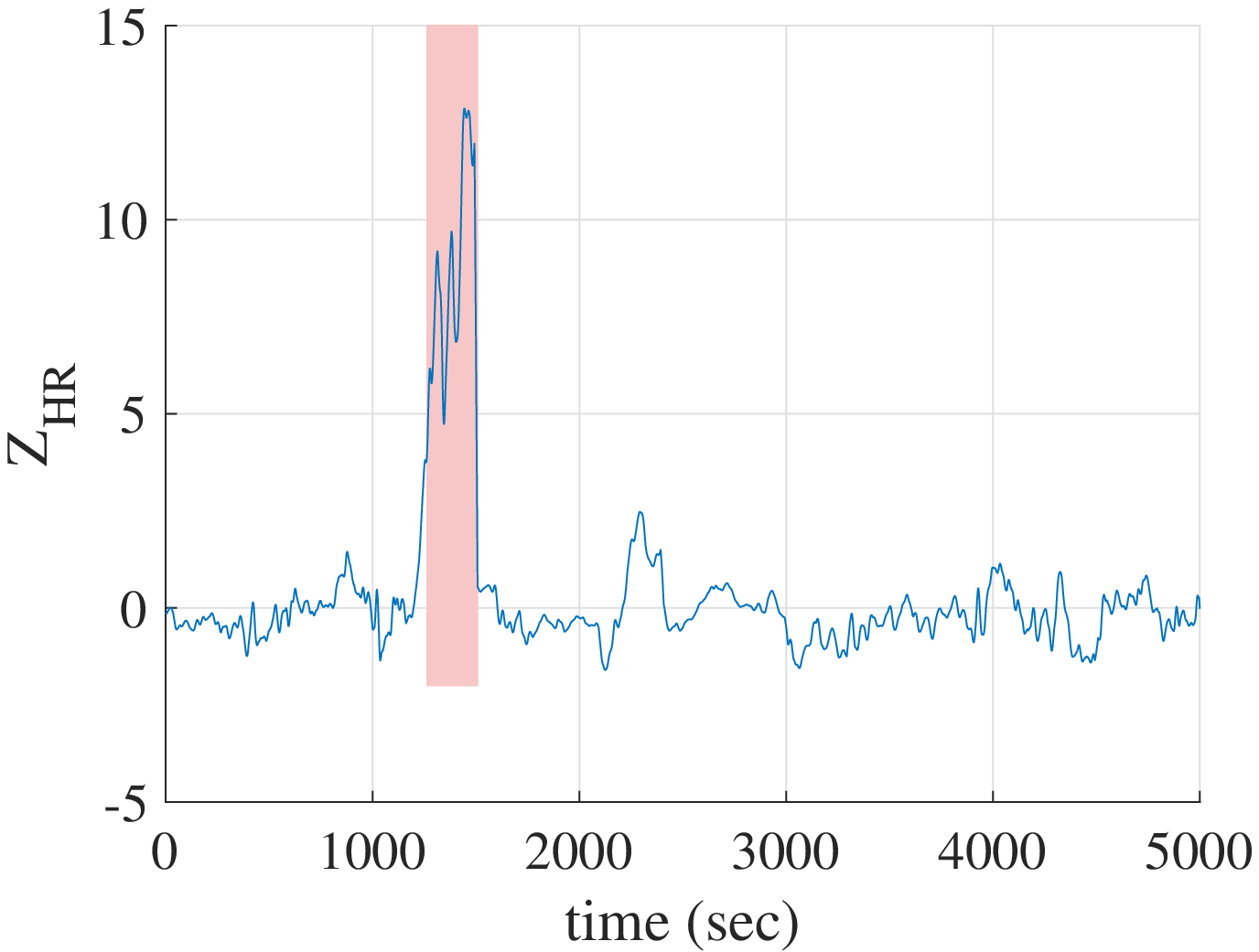}
        \caption{} 
	\end{subfigure}
\centering
	\begin{subfigure}{0.3\textwidth} 
		\includegraphics[trim=0 10 0 20,width=\textwidth]{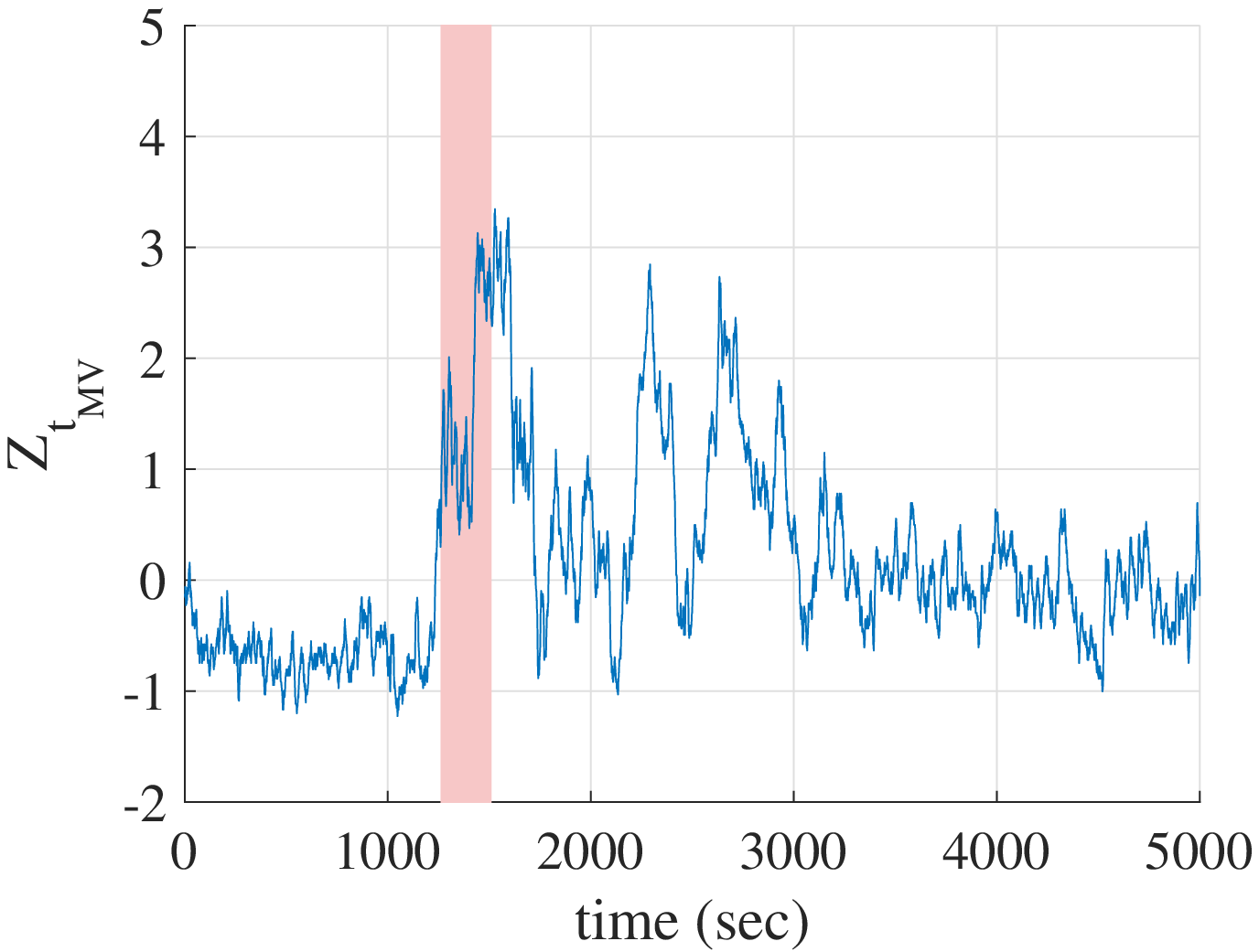}
        \caption{} 
	\end{subfigure}
\centering
	\begin{subfigure}{0.3\textwidth} 
		\includegraphics[trim=0 10 0 20,width=\textwidth]{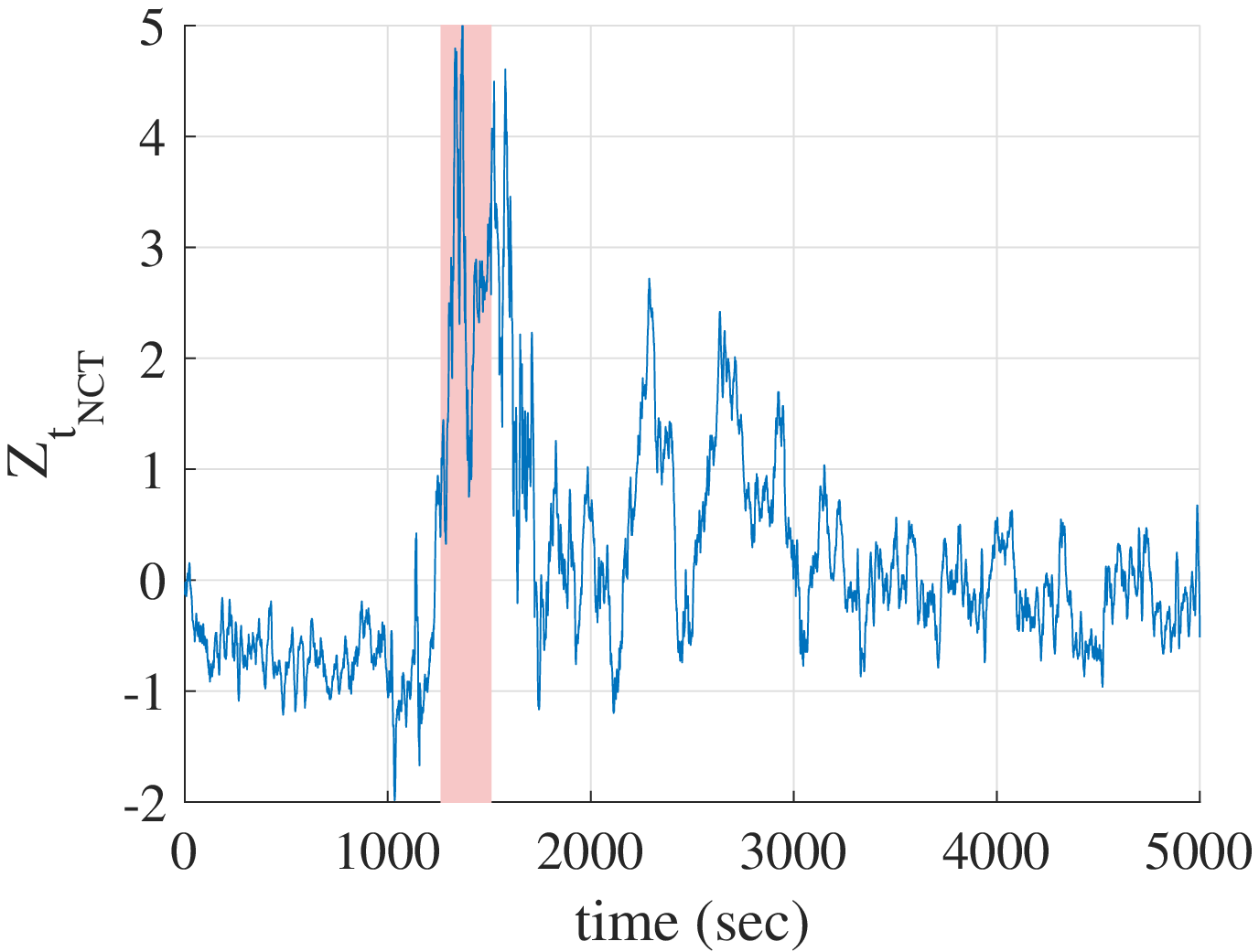}
        \caption{} 
	\end{subfigure}
\centering
	\begin{subfigure}{0.3\textwidth} 
		\includegraphics[trim=0 10 0 0,width=\textwidth]{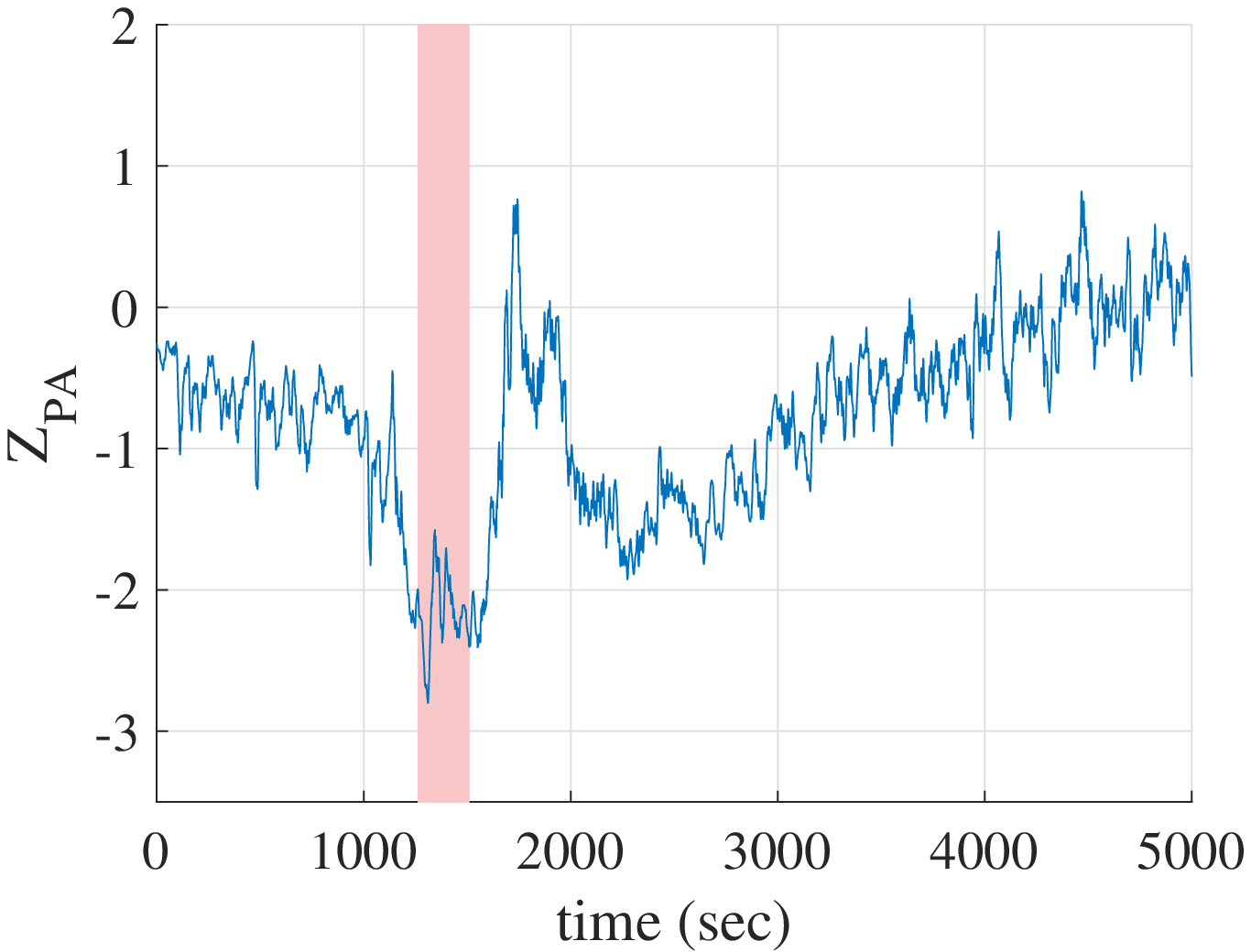}
        \caption{} 
	\end{subfigure}
\centering
	\begin{subfigure}{0.3\textwidth} 
		\includegraphics[trim=0 10 0 0,width=\textwidth]{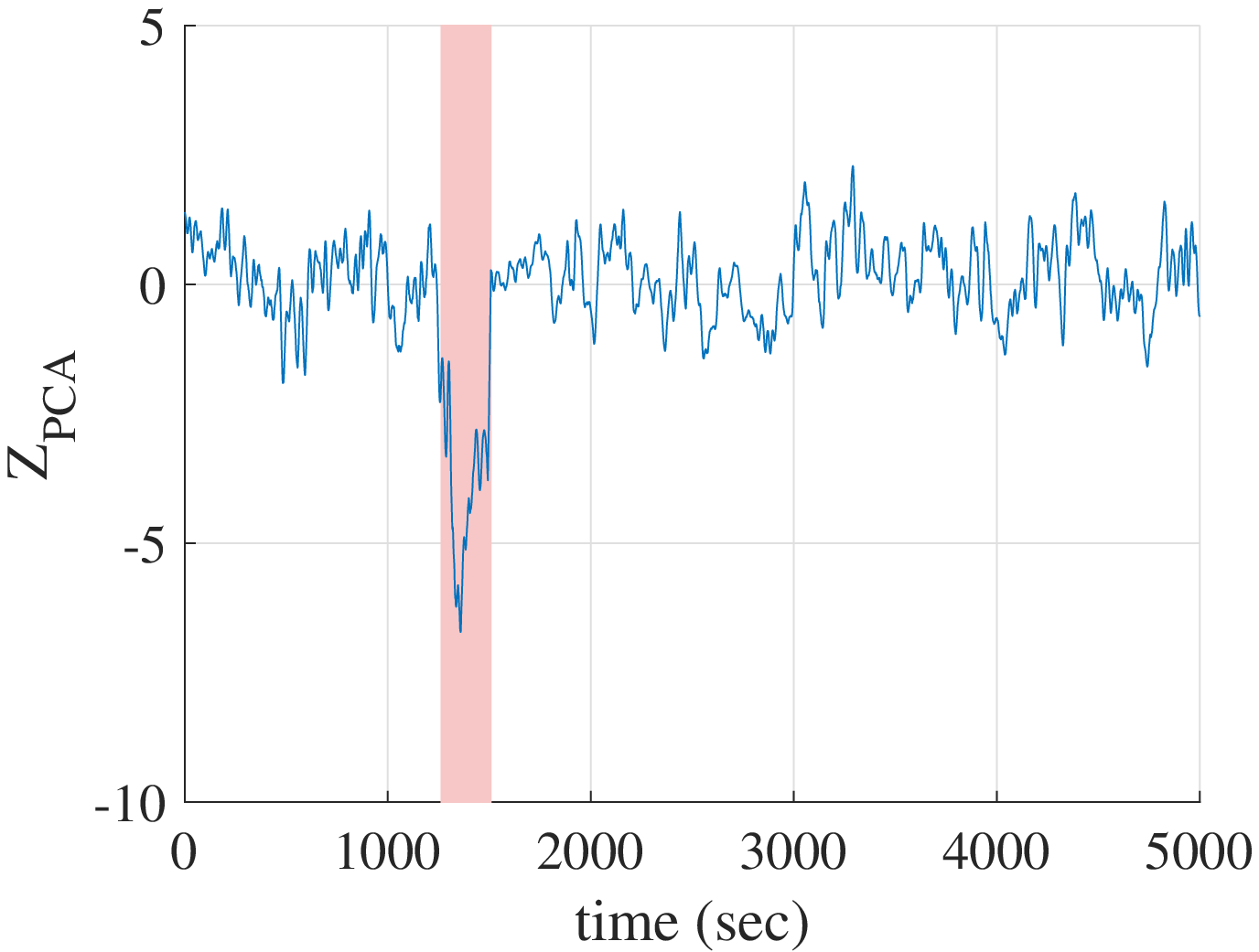}
        \caption{} 
	\end{subfigure}
\centering
	\begin{subfigure}{0.3\textwidth} 
		\includegraphics[trim=0 10 0 0,width=\textwidth]{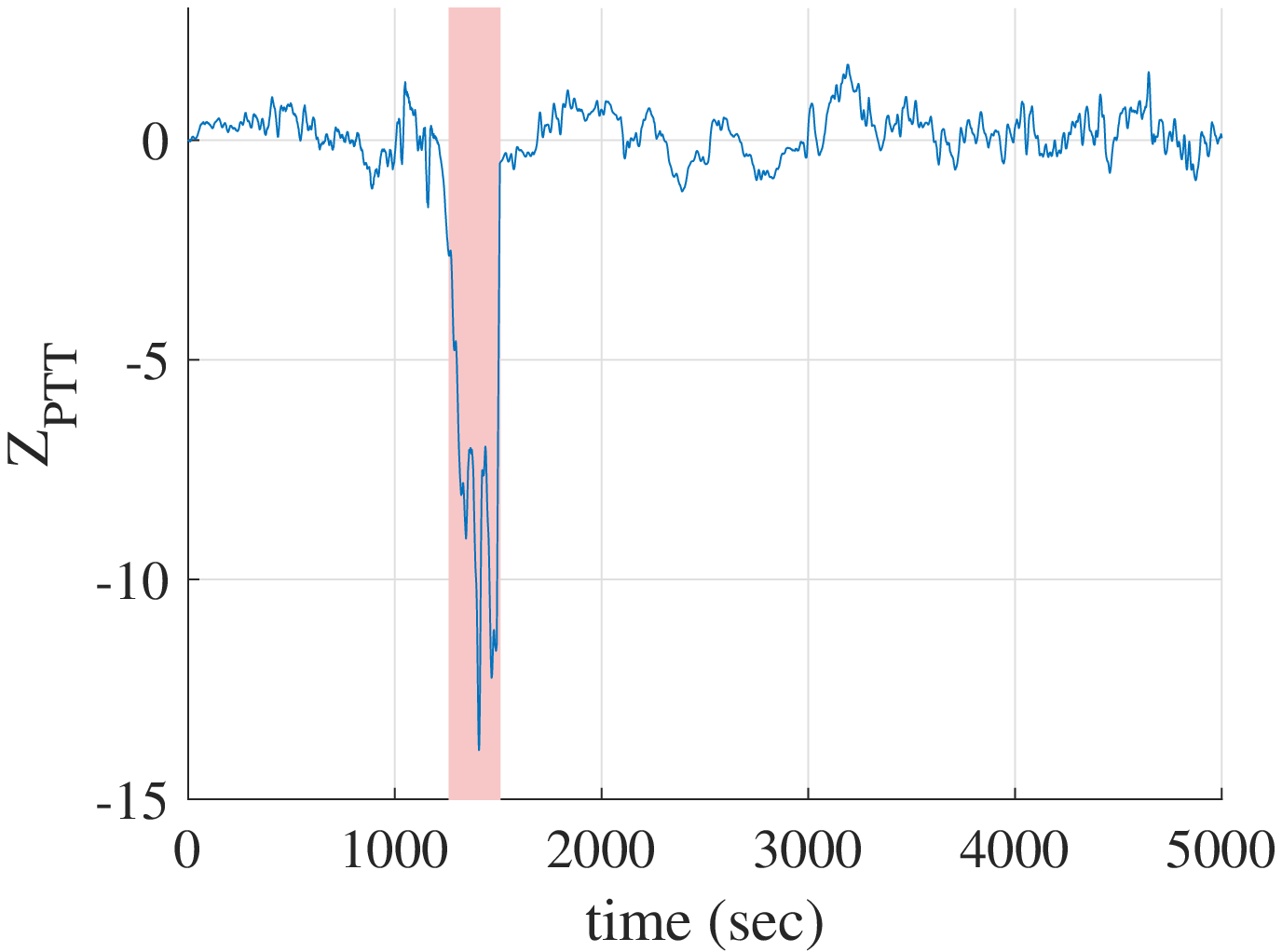}
        \caption{} 
	\end{subfigure}

\caption{Significant changes in z-scores proceeded by epileptic seizure in the first
	recorded seizure from Patient 6. (a) $z_{\text{HR}}$,(b) $z_{t_{\text{MV}}}$, (c)
	$z_{t_{\text{NCT}}}$, (d) $z_{\text{PA}}$,(e) $z_{\text{PCA1}}$, (f)
	$z_{\text{PTT}}$.}
\label{fig:onesiez}
\end{figure*}
\subsection{Preprocessing}
\label{subsec:preprocessing}
After synchronizing the PPG and EEG data, a neurologist marked the seizure onset and offset times based on the EEG waveform and video recordings. A total of 60 seizures were recognized among 12 subjects. Next, both PPG and EEG data were clipped so that it includes the seizure incidents as well as 2 hours of baseline before the seizure onset and 2 hours of data after the seizure offset. Thus, there is at least one seizure occurring in every 4 hours of the data. In case there are multiple seizures occurring close to each other(within less than 2 hours), the data is clipped such that it includes 2 hours of baseline before the first seizure and 2 hours of baseline after the last seizure. The clipped data was preprocessed to reject motion artifact and used for analysis in this study.
\subsubsection{Artifact Rejection}
\label{subsub:artifactRejection}
The PPG waveform is susceptible to motion artifact and can be completely obscured by noise. In order to reject noisy PPG pulses, the criteria introduced in \cite{segment_artifact} is employed. A clean PPG pulse should consist of a monotonically increasingly systolic upstroke followed by a diastolic descent. According to \cite{segment_artifact} the diastolic fall-off in a clean PPG pulse should not include more than two distinct notches. The PPG pulses with two, one, or zero dicrotic notches are considered clean. The artifactual parts are rejected in two phases. First, the algorithm proposed in \cite{segment_artifact}
	automatically marks the PPG troughs and detects the artifacts. In the second phase, an expert visually inspects the PPG data and manually removes any missed artifact and rejects the false positives. After the noise rejection, two seizures from subject 8 and one seizure from subject 11 were eliminated due to excessive amount of noise.
Table \ref{table:subjects} shows the duration of the clipped data for each specific subject. Table \ref{table:seizures} demonstrates the duration of the seizures being used for analysis. A total of 57 seizures were used for analysis.
\subsubsection{Feature Analysis}
\begin{figure*}
	\centering

	\begin{subfigure}{0.4\textwidth} 
		\includegraphics[trim=0 0 0 20,width=\textwidth]{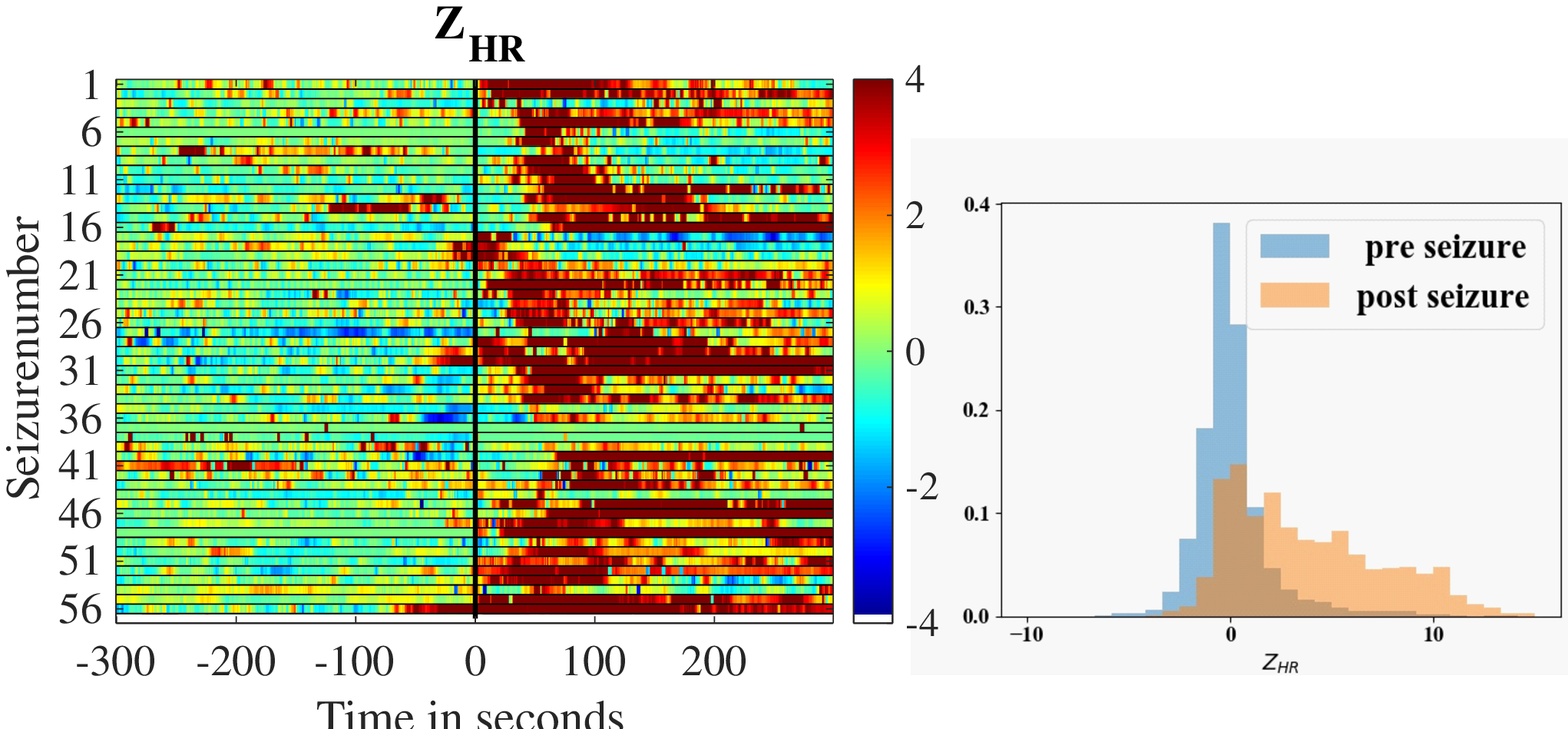}
        \caption{} 
	\end{subfigure}
    \begin{subfigure}{0.4\textwidth} 
		\includegraphics[trim=0 0 0 20,width=\textwidth]{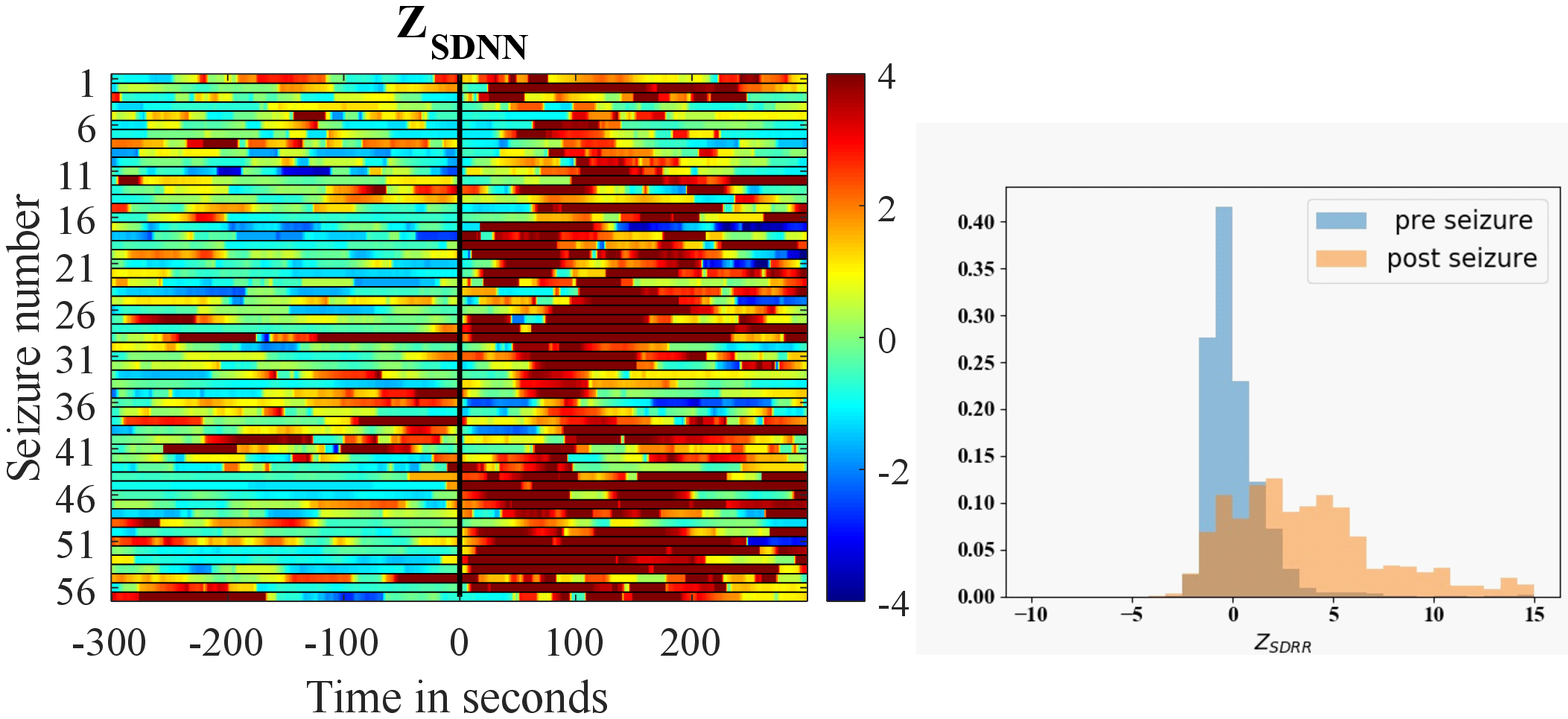}
        \caption{} 
	\end{subfigure}
    \begin{subfigure}{0.4\textwidth} 
		\includegraphics[trim=0 0 0 0,width=\textwidth]{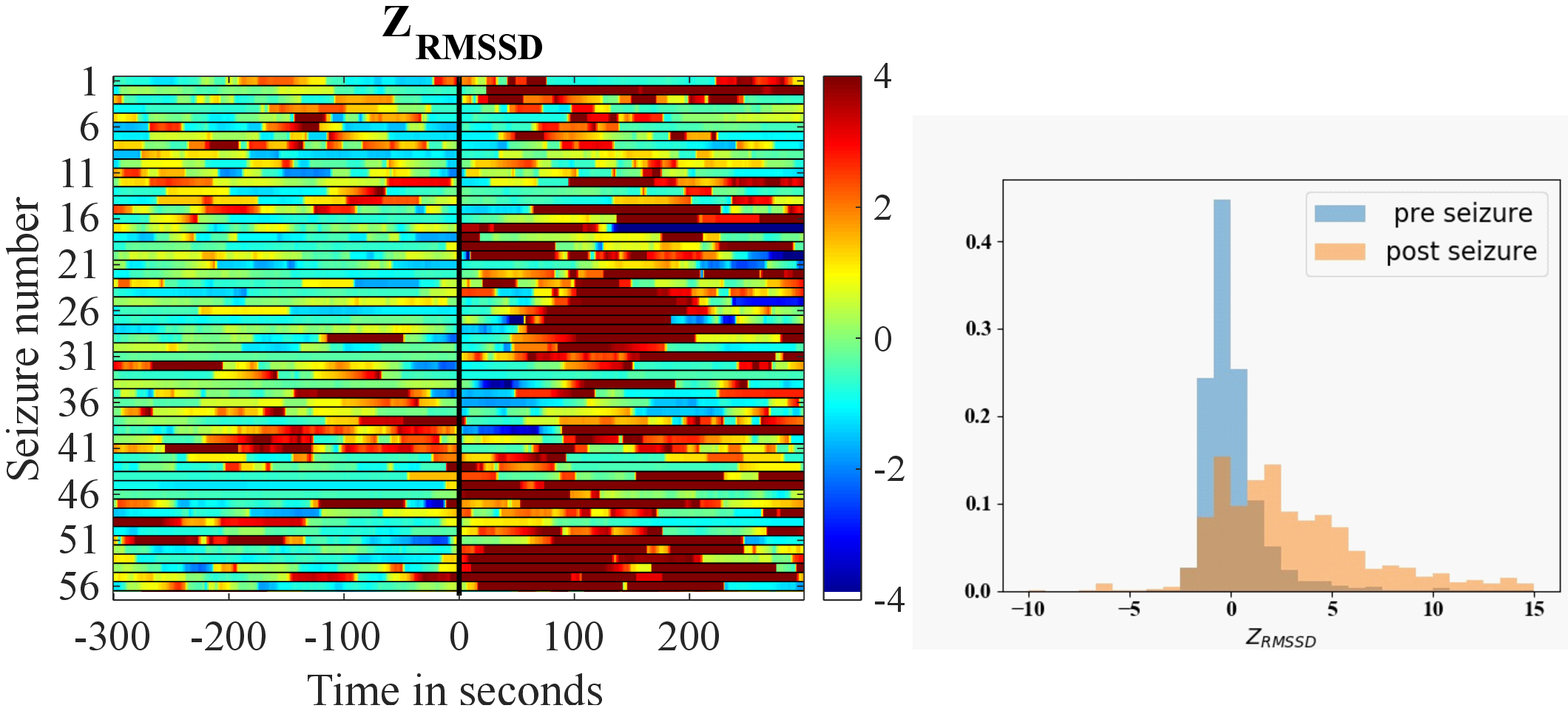}
        \caption{} 
	\end{subfigure}
    \begin{subfigure}{0.4\textwidth} 
		\includegraphics[trim=0 0 0 0,width=\textwidth]{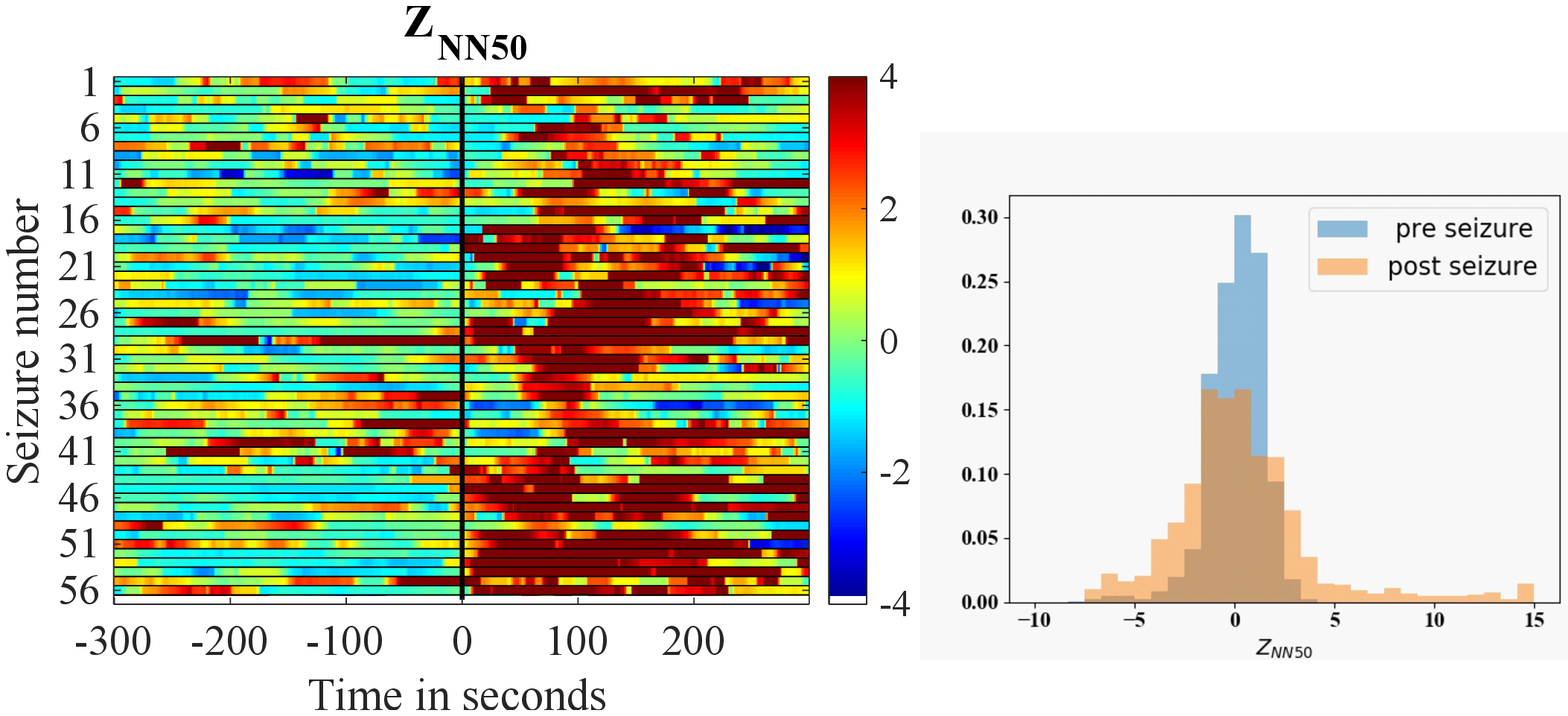}
        \caption{} 
	\end{subfigure}
    \begin{subfigure}{0.4\textwidth} 
		\includegraphics[trim=0 0 0 0,width=\textwidth]{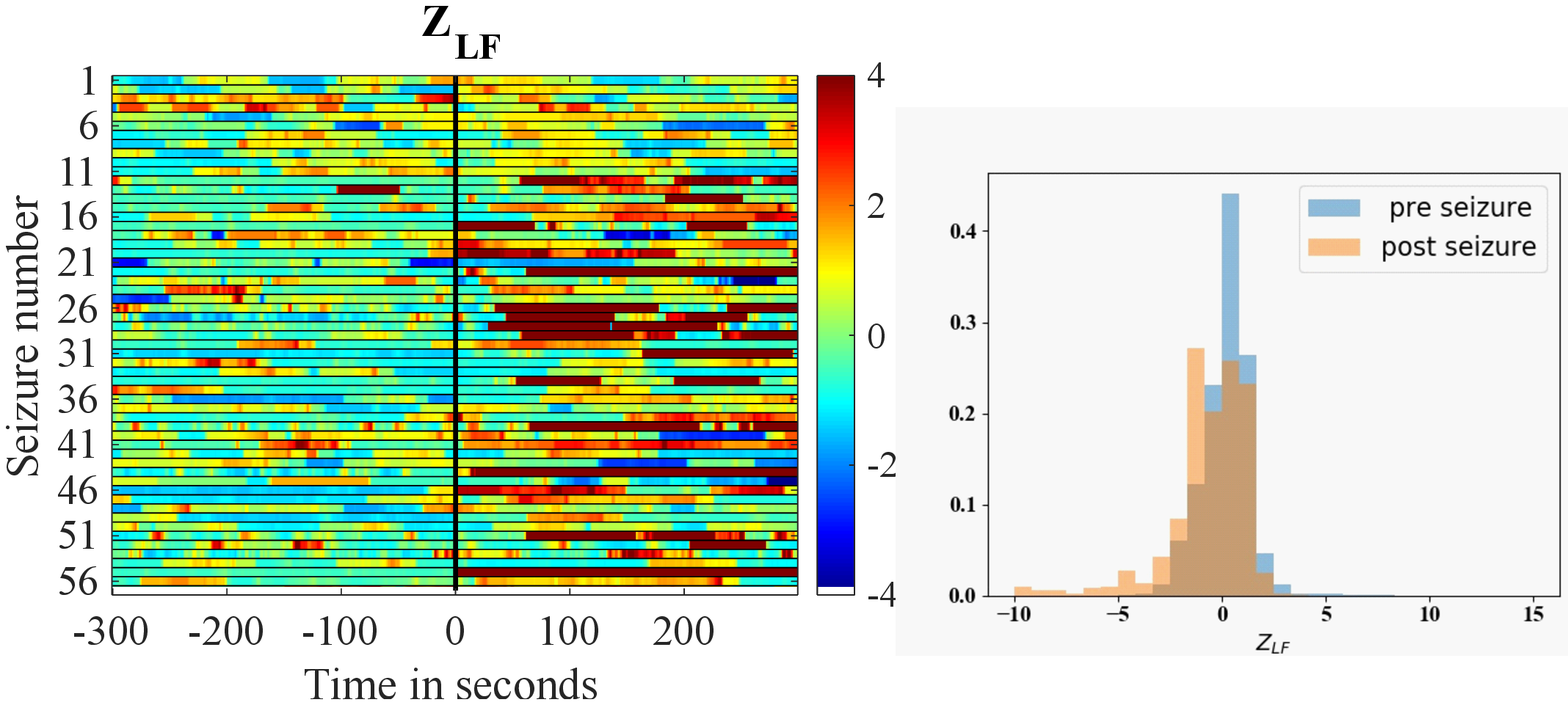}
        \caption{} 
	\end{subfigure}
    \begin{subfigure}{0.4\textwidth} 
		\includegraphics[trim=0 0 0 0,width=\textwidth]{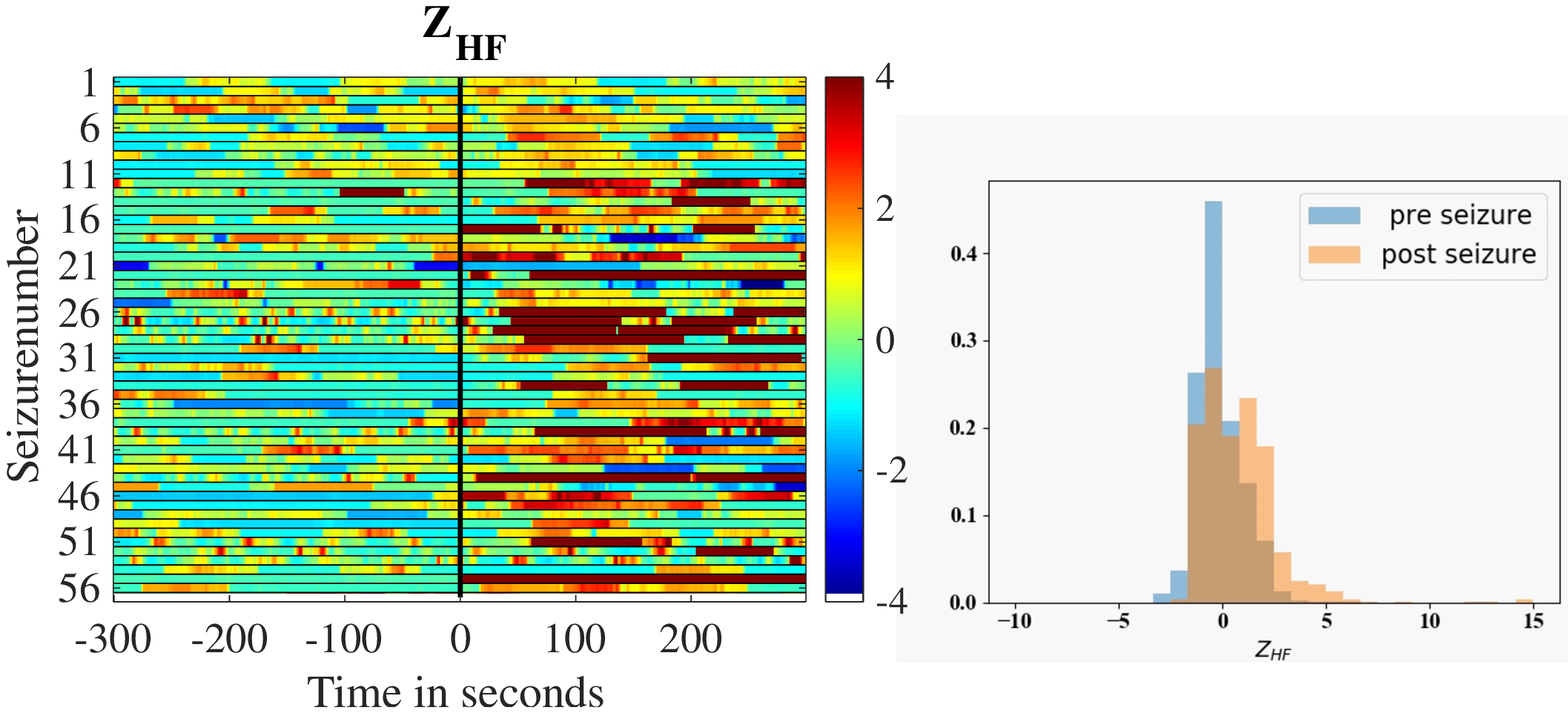}
        \caption{} 
	\end{subfigure}
    \begin{subfigure}{0.4\textwidth} 
		\includegraphics[trim=0 0 0 0,width=\textwidth]{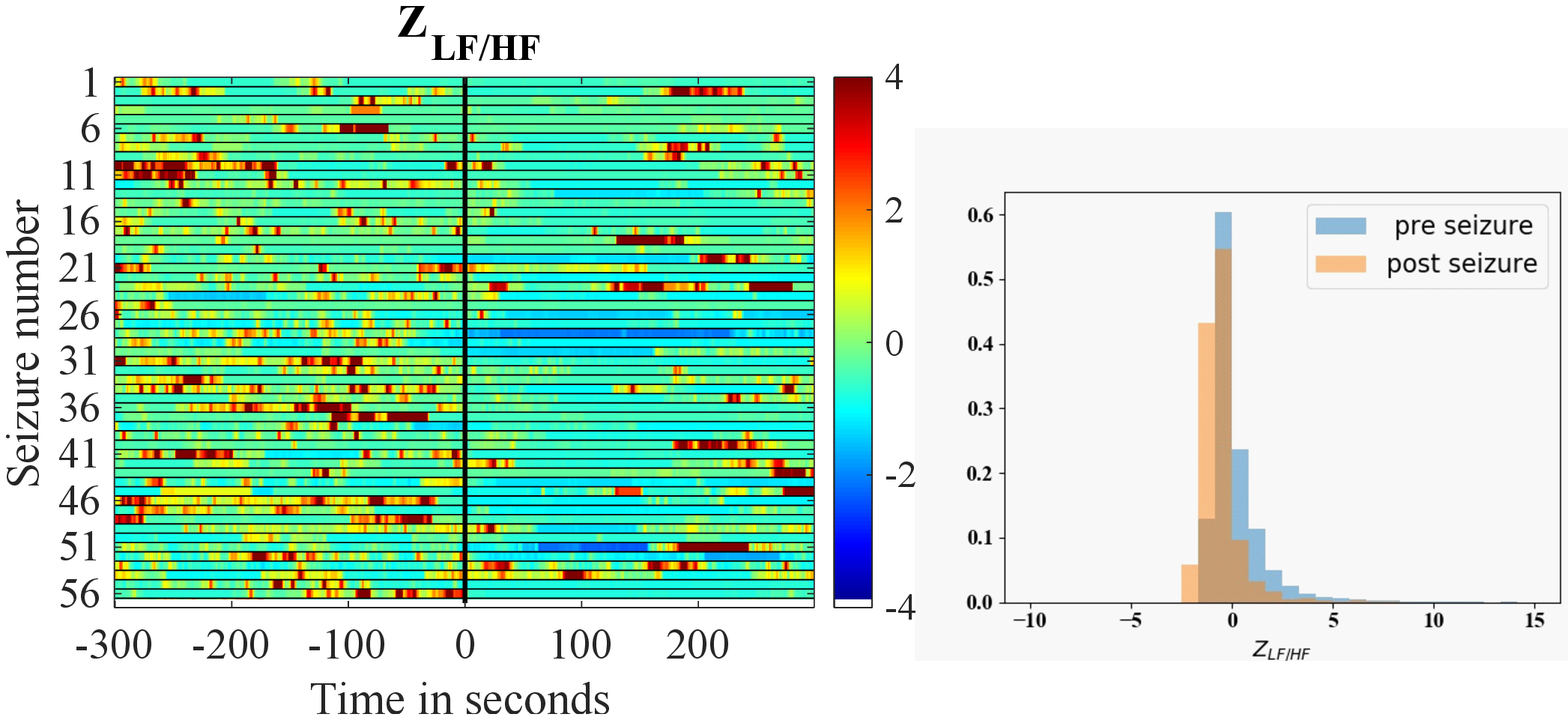}
        \caption{} 
	\end{subfigure}
	\centering
\begin{subfigure}{0.4\textwidth} 
		\includegraphics[trim=0 0 0 0,width=\textwidth]{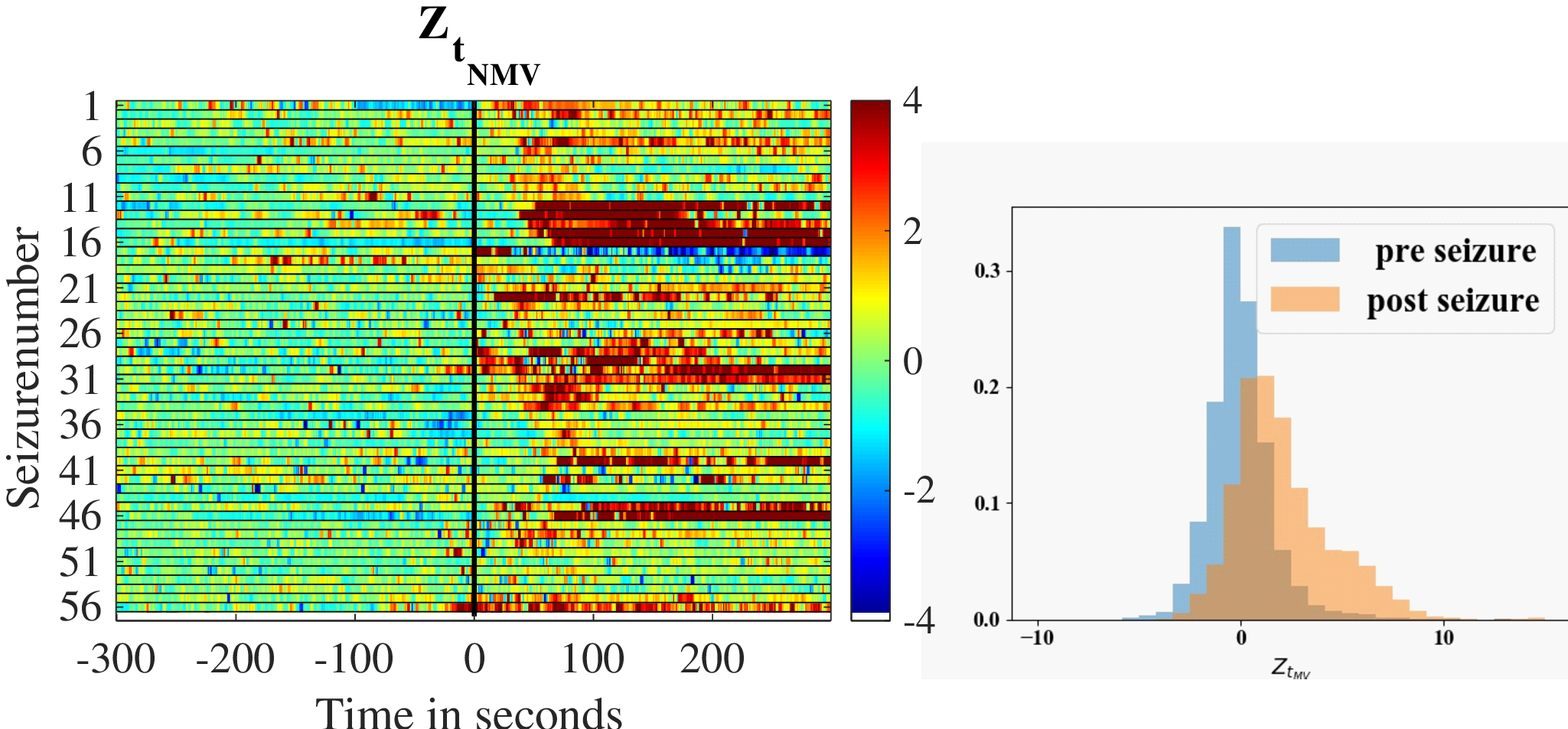}
        \caption{} 
	\end{subfigure}
    \begin{subfigure}{0.4\textwidth} 
		\includegraphics[trim=0 0 0 0,width=\textwidth]{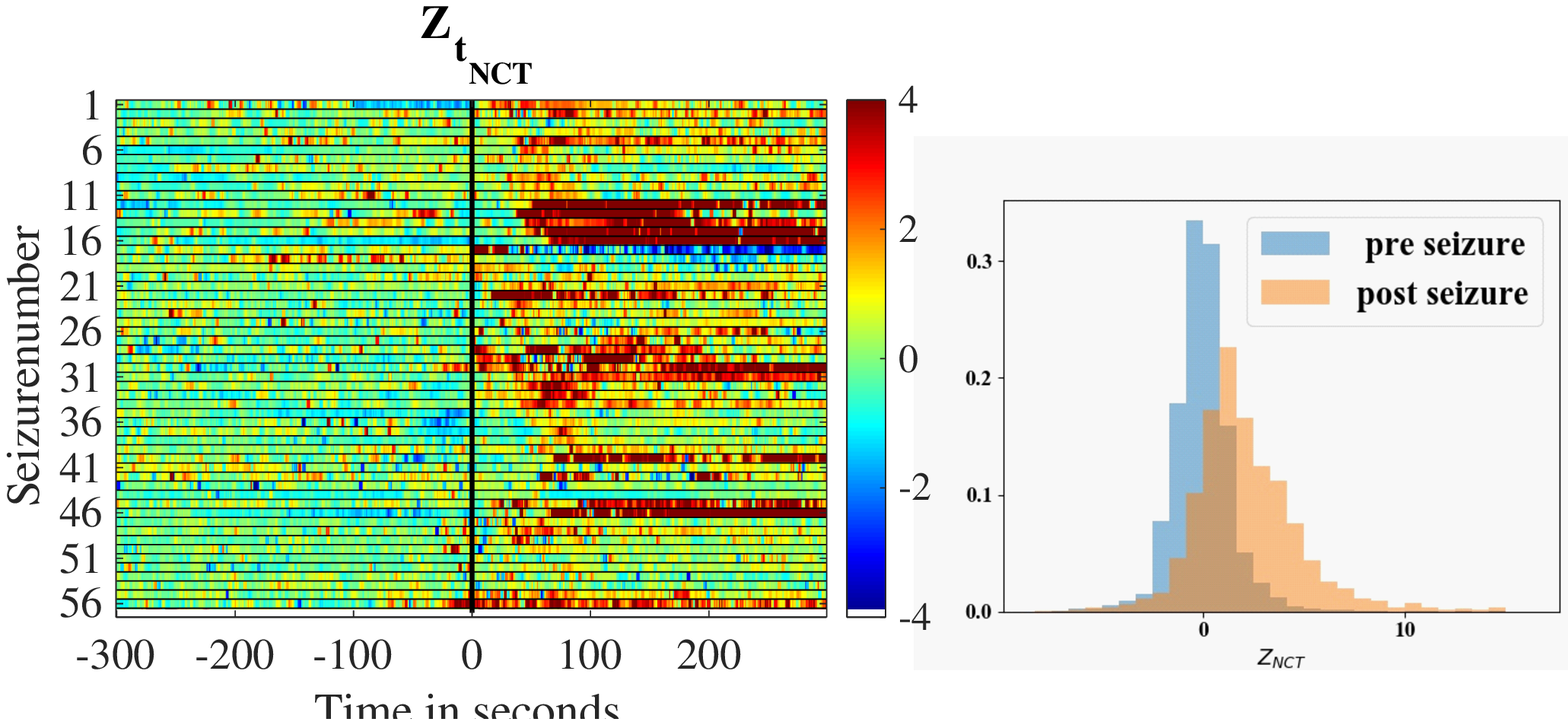}
        \caption{} 
	\end{subfigure}
    \begin{subfigure}{0.4\textwidth} 
		\includegraphics[trim=0 0 0 0,width=\textwidth]{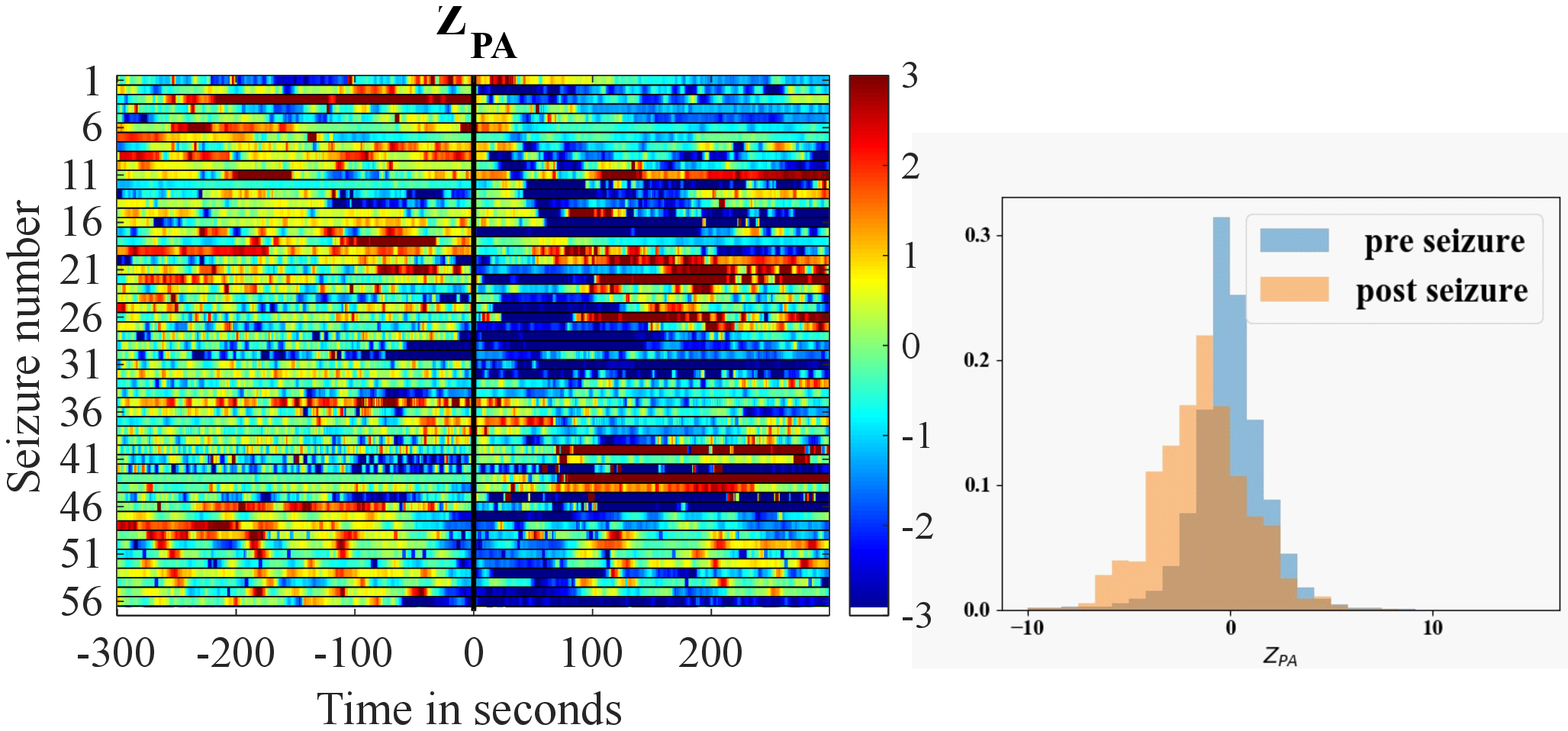}
        \caption{} 
	\end{subfigure}
    \begin{subfigure}{0.4\textwidth} 
		\includegraphics[trim=0 0 0 0,width=\textwidth]{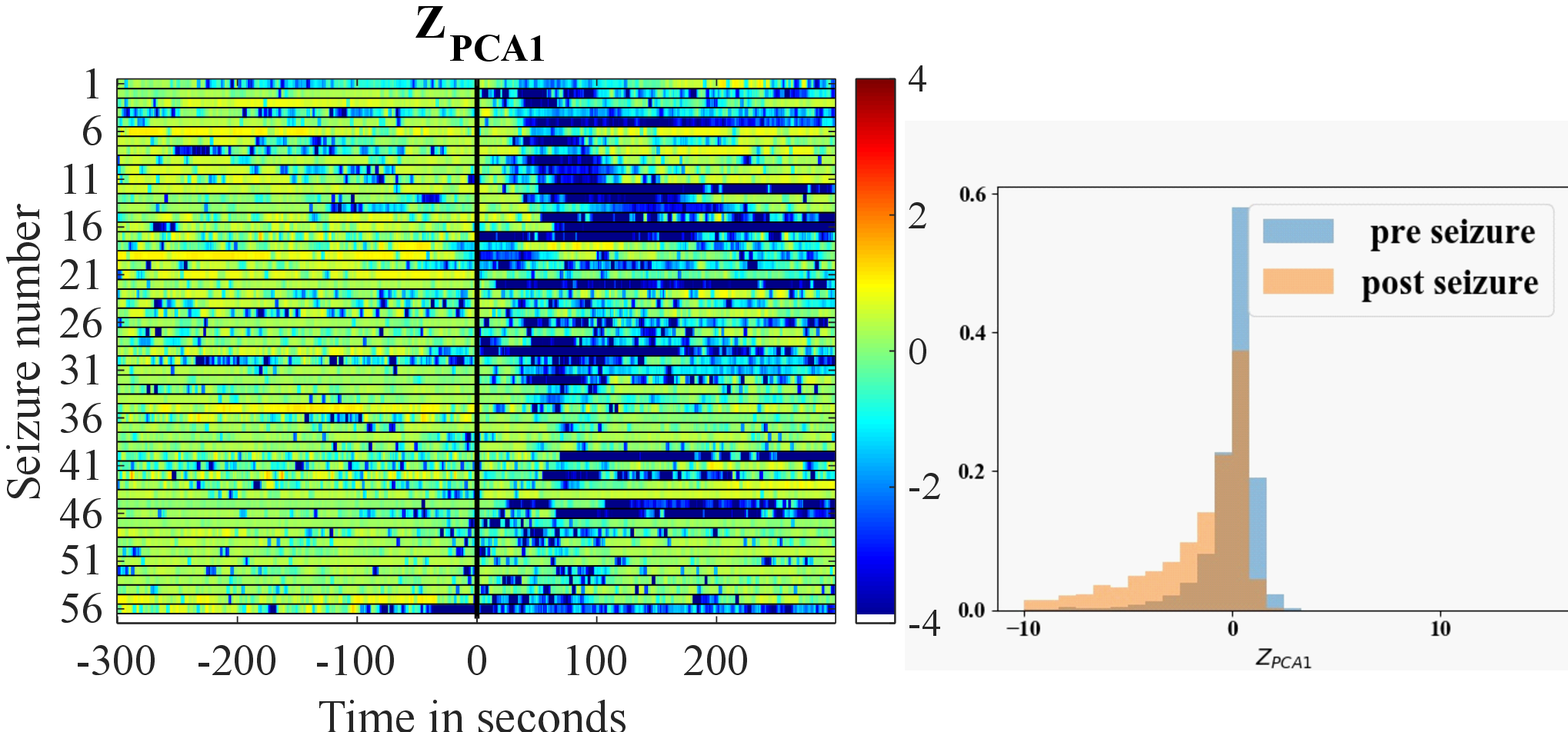}
        \caption{} 
	\end{subfigure}
    \begin{subfigure}{0.4\textwidth} 
		\includegraphics[trim=0 0 0 0,width=\textwidth]{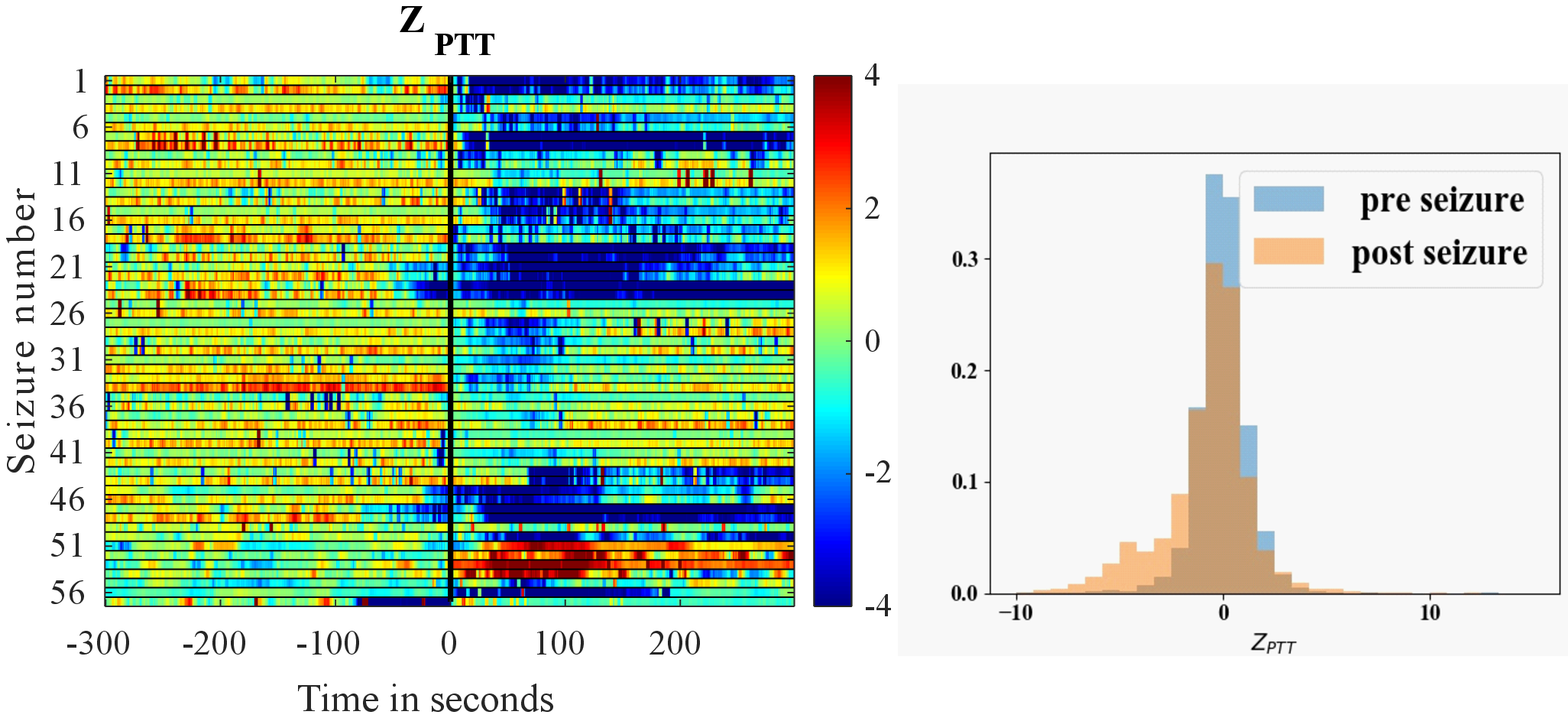}
        \caption{} 
	\end{subfigure}
\caption{The z-score values depicted in color for features derived from PPG signal. The x-axis represents the time for 5 minutes before and after the seizure onsets. The y-axis enumerates the seizures across all the subjects with total of 57 seizures. (a)$z_{HR}$, (b) $z_{SDNN}$, (c) $z_{RMSSD}$, (d) $z_{NN50}, $(e) $ z_{LF}$, (f) $z_{HF}$, (g)$z_{LF/HF}$, (h) $z_{t_{NMV}}$, (i) $z_{t_{NCT}}$, (j) $z_{PA}$, (k) $z_{PCA1}$, (l) $z_{PTT}$}.
\label{fig:hr_pdf}	
\end{figure*}

The morphological features were extracted from PPG data as defined in Section
\ref{sec:material}. Each feature forms a time series derived from the PPG pulses over time, where the individual samples belong to the feature values from each specific pulse. In order to compare the changes in
the features in time, each time series is segmented into non overlapping windows of 5-minutes length. Next,
the z-score of each feature sample is derived separately for every 5-min segment
based on the standard deviation and the mean value from its previous segment. For
instance, let us denote the mean and the standard deviation for the heart rate in
the $k^{th}$ segment  as $\mu_{(\text{HR})}^k$ and
$\sigma_{(\text{HR})}^k$. For one individual detected PPG pulse inside
$k^{\text{th}}$ segment, the heart rate is denoted as $\text{HR}^{k}[i]$ implying
that the $i^{\text{th}}$ pulse belongs to the $k^{\text{th}}$ segment. Next the z-score for HR feature in this pulse is derived as follows:

\begin{equation}\label{tvalue}
	z_{\text{HR}}^k[i]=\frac{\text{HR}^{k}[i]-\mu_{(\text{HR})}^{(k-1)}}{\sigma_{(\text{HR})}^{(k-1)}},
\end{equation}
where $\mu_{(\text{HR})}^{(k-1)}$ and $\sigma_{(\text{HR})}^{(k-1)}$ are the mean
and standard deviation from the previous segment. Normalizing the feature values based on the previous mean and standard deviation signifies the changes of these parameters over time. Let us denote the z-scores derived accordingly for all the hemodynamic-related features by $z_{\text{PA}}^{k}[i]$, $z_{t_{\text{NMVT}}}^{k}[i]$,
$z_{t_{\text{NCT}}}^{k}[i]$, $z_{\text{PCA1}}^{k}[i]$, and $z_{\text{PTT}}^{k}[i]$. \FIG{fig:onesiez} depicts  the variations of
z-scores derived for the hemodynamic-related features in the first seizure of subject 6. According to \FIG{fig:onesiez}, the heart rate, normalized crest time, and normalized maximum velocity time experience a significant increment after the seizure onsets, while the pulse amplitude, pulse transmit time, and the first principle coefficient encounter significant decrements. In order to be able to see the patterns of changes in all the recorded seizures, \FIG{fig:hr_pdf} demonstrates the z-scores derived for all the features.
 Comparing the z-scores for all the features pre- and post seizure shows a consistent pattern of changes across all the seizures.  In \FIG{fig:hr_pdf}, the middle black lines placed at zero time is when the seizure onsets happen. The negative
time values on x-axis are the 5-minute segments prior to seizures and the positive time
values demonstrate the 5-minute post seizure onset segments. The y-axis
shows the seizure numbers and the color codes represent z-score values between
$-4$ and 4, where the highlighted parts mean a higher z-score. In addition, the probability
distribution functions (pdf) of the derived z-scores for pre- and post-seizures are also depicted. The
blue diagram represents the z-scores derived using the data in the 5-minute
pre-seizure periods for all the 57 seizures. The orange diagrams depict the pdfs
of z-scores for all 57 post seizure onset features. In order to quantify the significance of the variations, we adopted one-way Analysis of variance (ANOVA) with the z-scores from interictal as the baseline and a 5-minutes of data after the seizure onset. Table \ref{table:sig} shows the percentage of seizures with significant changes after the seizure onset as well as the percentages of significant increases and decreases.


\begin{table*}[t]
\caption{Percentage of total recorded seizures showing significant changes ($p<0.01$) }
\centering
\begin{tabular}{c c c c c c c c c c c c c}
\hline
\hline
	Feature & HR &SDNN&RMSSD&NN50&LF&HF&LF/HF& $t_{\text{NMV}}$ & $t_{\text{NCT}}$ & \id{PA} & \id{PCA1}& \id{PTT} \\
  \hline
  Percentage of Significant Seizures   & 94.6\%&92.8\% &94.6\%&89.4\%&52.2\%&42.1\%& 52.2\%& 84.2\%& 84.2\% & 87.7\%& 66.6\% & 68.7\% \\
  Percentage of Significant Increase & 94.6\% &92.8\%&94.6\%&89.4\%&50.8\%&42.1\%& 0\%& 84.2\%& 84.2\% & 0\% & 0\% &8.7\%\\
  Percentage of Significant Decrease & 0\% &0\%&0\%&0\%& 1.4\%\%&0\% & 52.2\%&0\%&0\% & 87.7\% & 66.6\% &60\%\\

  \hline
  \hline \\

\end{tabular}
\label{table:sig}
\end{table*}
%
%
%
\section{Seizure Detection}
\label{sec:ML}
The entire data set consists of 102 hours and 30 minutes of recording among which a total of 1 hour and 9 minutes is in ictal phase. In order to quantify the performance, we adopted three metrics of sensitivity, Positive Predictive Value (PPV), and False Alarm Rate (FAR).
The sensitivity is defined as:
\begin{equation}\label{sens}
	\id{Sensitivity}=\frac{\id{TP}}{\id{TP}+\id{FN}},
\end{equation}
where $\id{TP}$ and $\id{FN}$ are true positives and false negatives, respectively. The FAR is the number of erroneously detected seizures per hour. The PPV is the ratio of true positives over all the issued alarms and is defined as:
\begin{equation}\label{ppv}
	\id{PPV}=\frac{\id{TP}}{\id{TP}+\id{FP}},
\end{equation}
where $\id{FP}$ is the false positives rate.
 In order to avoid overfitting, we randomly excluded a continuous segment of 2-hours inside the total of 102 hours 30 minutes data as the testing data. The remaining data is used for training the model. This random selection was repeated 20 times and the model was trained using the obtained training set. The overall performance is obtained by averaging over the sensitivity, PPV, and FAR from every repetition. We adopted a two layer LSTM neural network architecture. LSTM is able to capture long-term dependencies which helps identifying the temporal progression of PPG features in epilepsy.
\subsection{LSTM Architecture}
\begin{figure}[!t]
\begin{subfigure}{0.5\textwidth} 
		\includegraphics[trim=0 0 0 0,width=220pt]{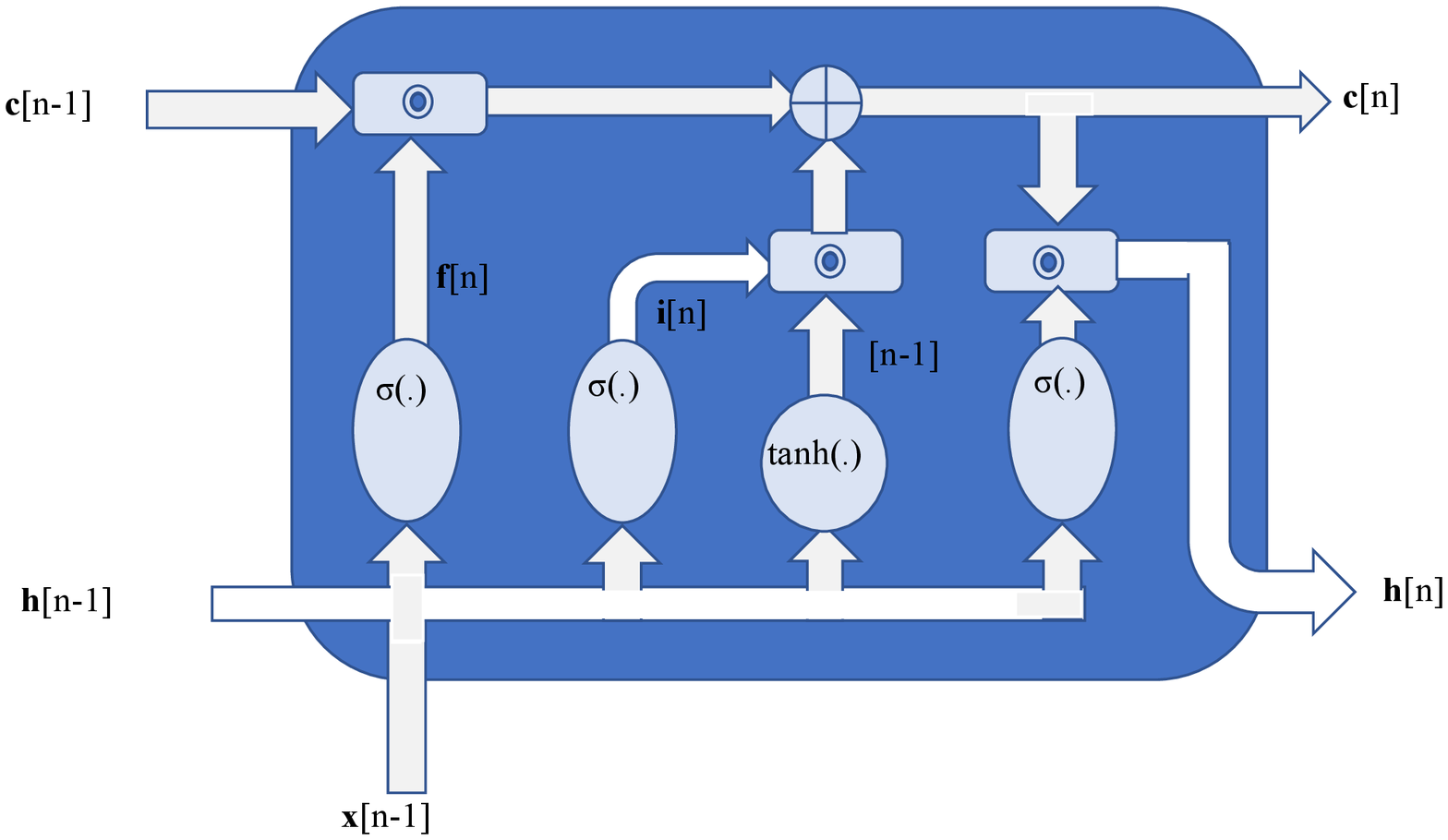}
        \caption{} 
	\end{subfigure}
\begin{subfigure}{0.45\textwidth} 
		\includegraphics[trim=0 0 0 0,width=220pt]{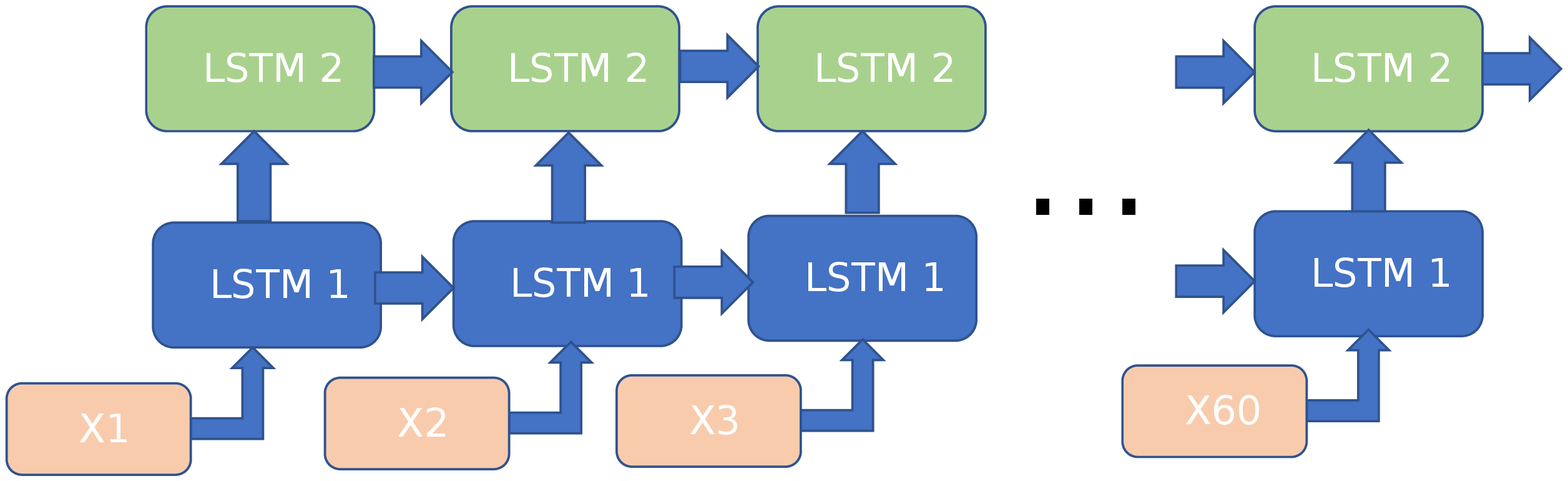}
        \caption{} 
	\end{subfigure}

\caption{The LSTM architecture.}
\label{fig:LSTM}
\end{figure}

The goal of automatic seizure detector is to classify windows of input data into
two labels of seizure and non-seizure. The stream of z-scores for each feature
forms a time series. A sliding window method was employed to subsegment the time
series into windows of 60 samples. \FIG{fig:LSTM} shows
the proposed architecture for the LSTM model. The LSTM architecture is composed of LSTM cells where each cell is fed with the hidden state and the cell state from the previous cells in time. Our proposed LSTM architecture contains two layers of  60 cells. A $20\%$ dropout was applied to the first layer to avoid over fitting. The Adam optimizer was used for training the architecture. \FIG{fig:LSTM} (a) shows the equations governing each cell, where $\text{h[n]}$ and $\text{c[n]}$ are the hidden state and the cell state corresponding to time $n$. $\boldsymbol{W}$ and $\boldsymbol{b}$ are the weight matrices and the bias vectors.  The hidden state is dependent on the input vector, the previous hidden state and the cell state. The vector $\boldsymbol{i}[n]$ decides whether to use the input and the past hidden state to update the cell state $\boldsymbol{c}[n]$. The vector $\boldsymbol{f}[n]$ decides wether to use the past cell state $\boldsymbol{c}[n-1]$ to update the cell state. The equations governing the LSTM cells are as follows:
\begin{align}\label{equ:lstm}
   &\boldsymbol{f}[n]=\sigma\big(\boldsymbol{W}_f\big[\boldsymbol{h}[n-1],\boldsymbol{x}[n]\big]+\boldsymbol{b}_f\big)\nonumber\\
   &\boldsymbol{i}[n]=\sigma\big(\boldsymbol{W}_i\big[\boldsymbol{h}[n-1],\boldsymbol{x}[n]\big]+\boldsymbol{b}_i\big)\nonumber\\
   &\boldsymbol{\tilde{c}}[n]=tanh\big(\boldsymbol{W}_c\big[\boldsymbol{h}[n-1],\boldsymbol{x}[n]\big]+\boldsymbol{b}_c\big)\nonumber\\
   &\boldsymbol{o}[n]=\sigma\big(\boldsymbol{W}_o\big[\boldsymbol{h}[n-1],\boldsymbol{x}[n]\big]+\boldsymbol{b}_o\big)\nonumber\\
   &\boldsymbol{c}[n]=\boldsymbol{f}[n]\odot \boldsymbol{c}[n-1]+ \boldsymbol{i}[n]\odot \tilde{\boldsymbol{c}[n]}\nonumber\\
   &\boldsymbol{h}[n]=\boldsymbol{o}[n]\odot tanh\big(\boldsymbol{c}[n]\big),
\end{align}
where $\odot$ is the element-wise multiplication. \\
\subsection{HRV-related versus hemodynamic-related features}
\indent Conventional seizure detection algorithms, derive the HRV-based features from ECG signal and use them for seizure detection. In this work, the HRV-related features were extracted from PPG signal, i.e. $\text{HR}$, $\text{SDNN}$, $\text{RMSSD}$, $\text{NN50}$, $\text{LF}_{norm}$, $\text{HF}_{norm}$, and $\text{LF/HF}$. In addition to the HRV-related features, 5 other PPG features were studied  which are known to be related to vascular compliance and hemodynamics, i.e. $t_{NMV}$, $t_{NCT}$, $t_{PA}$, $PCA1$, $PTT$. As it Despite the HRV-related features, all the hemodynamic-related features show consistent patterns of ictal change across majority of the seizures. In order to be able to measure how these hemodynamic-related features are contributing to the improvement of seizure detection, we trained the proposed LSTM architecture twice. First using the 7 HRV-related features,  the dimensions of the input vector and the hidden state and cell state for the first layer were 7. The dimensions of the hidden state and the cell state vectors for the second layer were chosen to be 5. The performance results of the trained HRV-based LSTM is brought in table \ref{table:result}. For brevity, this detector is called LSTM7.\\
\indent Next, the 5 hemodynamic-related features were appended to the input vector, making the input vectors of 12 features in time. The hemodynamic-based LSTM is denoted by LSTM12 in table \ref{table:result}.T he dimensions of the hidden states and the cell states for the first and the second layer were 12 and 5 respectively. As shown in table \ref{table:result}, although our proposed seizure detector is subject independent, our HRV-based detector LSTM7 is showing an improvement in FAR and PPV compared to \cite{ECG_feedback,ECG_singlelead}. The LSTM7 results performance is comparable to \cite{ECG_nocturnal}, considering the fact that the data in \cite{ECG_nocturnal} is nocturnal, while our data is both nocturnal and diurnal. Adding the 5 hemodynamic-related features in LSTM12 improves the sensitivity to $92\%$. In addition , the FAR rate have improved from 0.91 to 0.52 which is a $\text{42}\%$ improvement.
\begin{table*}[t]
\caption{Results}
\centering
\begin{tabular}{c c c c c c c}
\hline
\hline
	work & sensitivity & PPV &FAR  & type of seizure & time & subject dependent \\
  \hline
   \cite{ECG_nocturnal} & 77.6\%\% & 30.7\%& 0.33 per hour & GTC + partial& nocturnal & Y \\
  \cite{ECG_singlelead}& 81.9\% & 7.9\%&1.97 per hour & GTC+partial & nocturnal+diurnal &Y\\
   \cite{ECG_feedback}& 77.1\% & 3.25\%&1.24 per hour & GTC+partial & nocturnal+diurnal &Y\\
   \cite{MPDIsensor}& 77.1\% & 3.25\%&1.24 per hour & GTC+partial & nocturnal+diurnal &Y\\
  LSTM based on HRV (LSTM7) &82\%&26\%&0.91 per hour&GTC+partial&nocturnal+dirunal&N\\
  LSTM based on PPG morphological &92\%&43\%&0.52 per hour&GTC+partial&nocturnal+dirunal&N\\
  features (LASTM12)&&&&&&\\

  \hline
  \hline \\

\end{tabular}
\label{table:result}
\end{table*}

%

\section{Conclusion}
\label{sec:conclusion}
The automatic seizure detectors using frequency domain features of EEG signals and traditional machine learning techniques such as k-nearest neighbor and Support Vector Machine (SVM) have been extensively researched in literature \cite{kNN,nonlinear} in the past. Recently, deep neural network techniques have gained attention enabling seizure prediction from pre-ictal EEG patterns with high sensitivity and low FAR \cite{LSTM}. Since the contribution of different channels in seizure detection or prediction performance is patient specific and depends on the seizure onset zone, these models cannot
be generally used for every one and require subject-specific sensor selection and optimization techniques to reduce the complexity \cite{sensor_selection}. In addition, wearing an EEG cap is not feasible for a long-term monitoring of patients leading a normal daily life. 

The autonomic nervous system dysfunction induced by epileptic brain activity is easier to detect using extracerebral sensors \cite{discharge}. In particular, heart rate variability (HRV), accelerometry and Electrodermal Activity (EDA) have gained attention for measuring the autonomic dysfunction. \cite{EMG,ACC}The frequency domain features such as Very Low Frequency (VLF)(0.0033-0.04~Hz), Low Frequency (LF)(0.04-0.15~Hz), and High Frequency (HF) (0.15-0.4~Hz) measure the Sympathetic (SNS) and parasympathetic (PNS) nervous balance \cite{heart-brain}. However, recent studies have suggested the LF/HF ratio is a controversial measure of sympatho/vagal balance. Since SNS and PNS activities are nor linearly counterparts, decreasing activity in one does not suggest an increase in the other. In fact, both SNS and PNS are contributing to LF power, whereas HF power is associated with PNS activity \cite{LF-HF}.

The Catecholamines released as anticonvulsant during epileptic seizures affect the
blood circulation and vascular function. The hemodynamics induced by seizures can
be measured by PPG sensors. In this study, we investigated six different
morphological features in PPG and ECG signal recorded from 12 epileptic subjects. A
total of 102 hours and 30 minutes of ictal and inter-ictal data was recorded. The
extracted features include heart rate, crest time, maximum velocity time, pulse
transmit time, pulse amplitude, and the first principle coefficient derived from
the second derivative of PPG pulse shape. Among these features, the crest time, the
maximum velocity time, and the pulse transmit time are influenced by the heart rate. In order to eliminate the effect of heart rate, these feature were normalized to pulse width.

All the investigated features showed significant variation in majority of recorded
seizures (refer to Table \ref{table:sig}). Except for patient number 11, who had a few seizures in which the pulse-transmit time was significantly increased,  a consistent pattern across all seizures/subjects was observed. The reduction in pulse amplitude and the increase in normalized crest time suggests an increase in vascular resistance and hypovolemia in limbs induced by vasoconstriction. This is the first time a comprehensive analysis on PPG morphology along with ECG data induced by epilepsy is performed. 

\end{document}